\documentclass[10pt,twocolumn]{article}

\usepackage[dvips]{graphicx}
\usepackage[english]{babel}
\usepackage{hyperref}
\usepackage{amsmath}

\allowdisplaybreaks[2]

\newcommand{\half}{{\textstyle\frac{1}{2}}}

\newcommand{\gev}{\operatorname{GeV}}

\newcommand{\ms}{\mskip 1.5mu}
\newcommand{\bs}{\mskip -1.5mu}

\newcommand{\tvec}[1]{\boldsymbol{#1}}

\newcommand{\mat}[1]{\langle\!\langle\ms #1
                     \ms\rangle\!\rangle}

\newcommand{\mvec}[1]{\smash{\vec{\mskip 0.5mu #1}}\mskip 1.5mu}

\newcommand{\sing}[1]{{}^{1\!}#1}
\newcommand{\oct}[1]{{}^{8\!}#1}
\newcommand{\octA}[1]{{}^{A\!}#1}
\newcommand{\octS}[1]{{}^{S\!}#1}

\begin{document}

\title{%
\raggedleft{\normalsize DESY 11-021} \\[1.0em]
\centering{\textbf{\Large Theoretical considerations on multiparton
  interactions in QCD}} \\
}

\author{Markus~Diehl \\[0.2em]
\textit{Deutsches Elektronen-Synchroton DESY, 22603 Hamburg, Germany}
\and
Andreas Sch\"afer \\[0.2em]
\textit{Institut f\"ur Theoretische Physik, Universit\"at Regensburg,
  93040 Regensburg, Germany}
}

\date{\parbox{0.9\textwidth}{\normalsize \quad\textbf{Abstract:} We
    investigate several ingredients for a theory of multiple hard
    scattering in hadron-hadron collisions.  Issues discussed include the
    space-time structure of multiple interactions, their power behavior,
    spin and color correlations, interference terms, scale evolution and
    Sudakov logarithms.  We discuss possibilities to constrain multiparton
    distributions by lattice calculations and by connecting them with
    generalized parton distributions.  We show that the behavior of
    two-parton distributions at small interparton distances leads to
    problems with ultraviolet divergences and with double counting, which
    requires modification of the presently available theoretical
    framework.
}}

\maketitle


\section{Introduction}

In hadron-hadron collisions at very high energies, several partons in one
hadron can scatter on partons in the other hadron and produce particles
with large transverse momentum or large invariant mass.  The effects of
such multiparton interactions average out in sufficiently inclusive
observables, which can be described by conventional factorization formulae
that involve a single hard scattering.  However, multiple interactions do
change the structure of the final state.  They have been seen at the
Tevatron \cite{Abe:1997bp,Abazov:2009gc} and are expected to be important
for many analyses at LHC
\cite{Jung:2009eq,Alekhin:2005dx,Bartalini:2010su}.

The phenomenology of multiparton interactions relies on models that are
physically intuitive but involve significant simplifications.  A brief
review of the subject can be found in \cite{Sjostrand:2004pf} and an
overview of implementations in event generators in \cite{Buckley:2011ms}.
So far a systematic description of multiple interactions in QCD remains
elusive.  In this letter we report on some steps towards this goal.  We
will see to which extent the cross section formulae currently used to
calculate multiple-scattering processes can be justified in QCD and to
which extent they need to be completed.  At a more fundamental level, we
find that there is an unsolved problem of double counting between single
and multiple hard scattering.

We consider the case of two hard scatters at parton level.  For
definiteness we analyze the production of two electroweak gauge bosons
with large invariant mass ($\gamma^*$, $Z$ or $W$) and indicate which of
our results can be generalized to other final states such as jets.  Since
the main interest in multiparton interactions is driven by the need to
understand details of the final state, we keep the transverse momenta of
the produced gauge bosons differential, rather than integrating over them.
For the production of a single boson there is a powerful theoretical
description based on transverse-momentum dependent parton densities
\cite{Collins:1981uk,Ji:2004wu,Collins:2007ph, Collins:2011}, which we aim
to extend to the case of multiparton interactions.  Integrating over
transverse momenta gives the more familiar formulation in terms of
collinear parton distributions.  Detailed derivations of our results and
further discussion will be given in \cite{Diehl:2011}.


\section{Tree-level analysis}
\label{sec:tree}

\begin{figure*}
\begin{center}
\hspace{1em}
\includegraphics[height=0.28\textwidth]{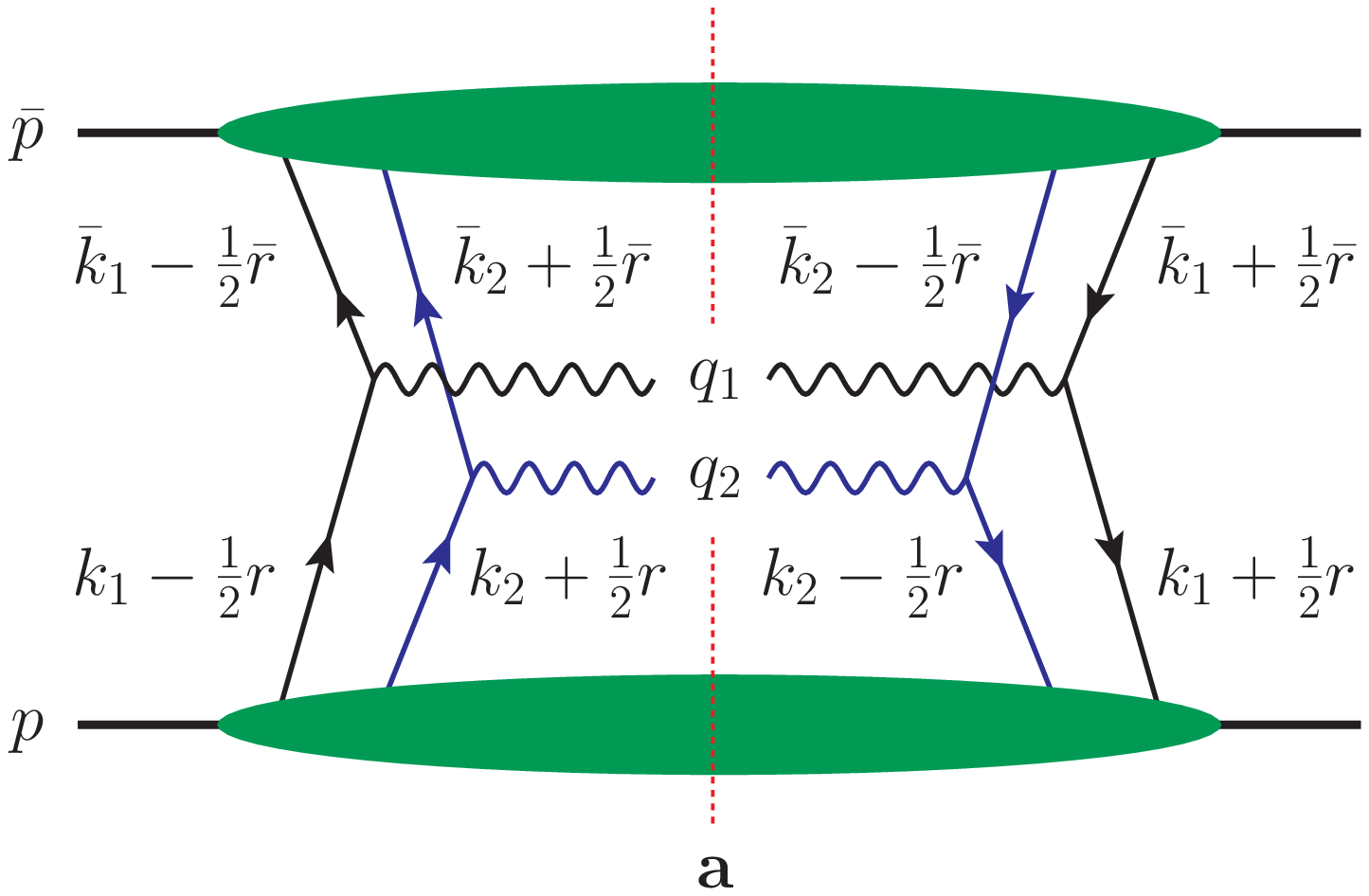}
\hfill
\includegraphics[height=0.28\textwidth]{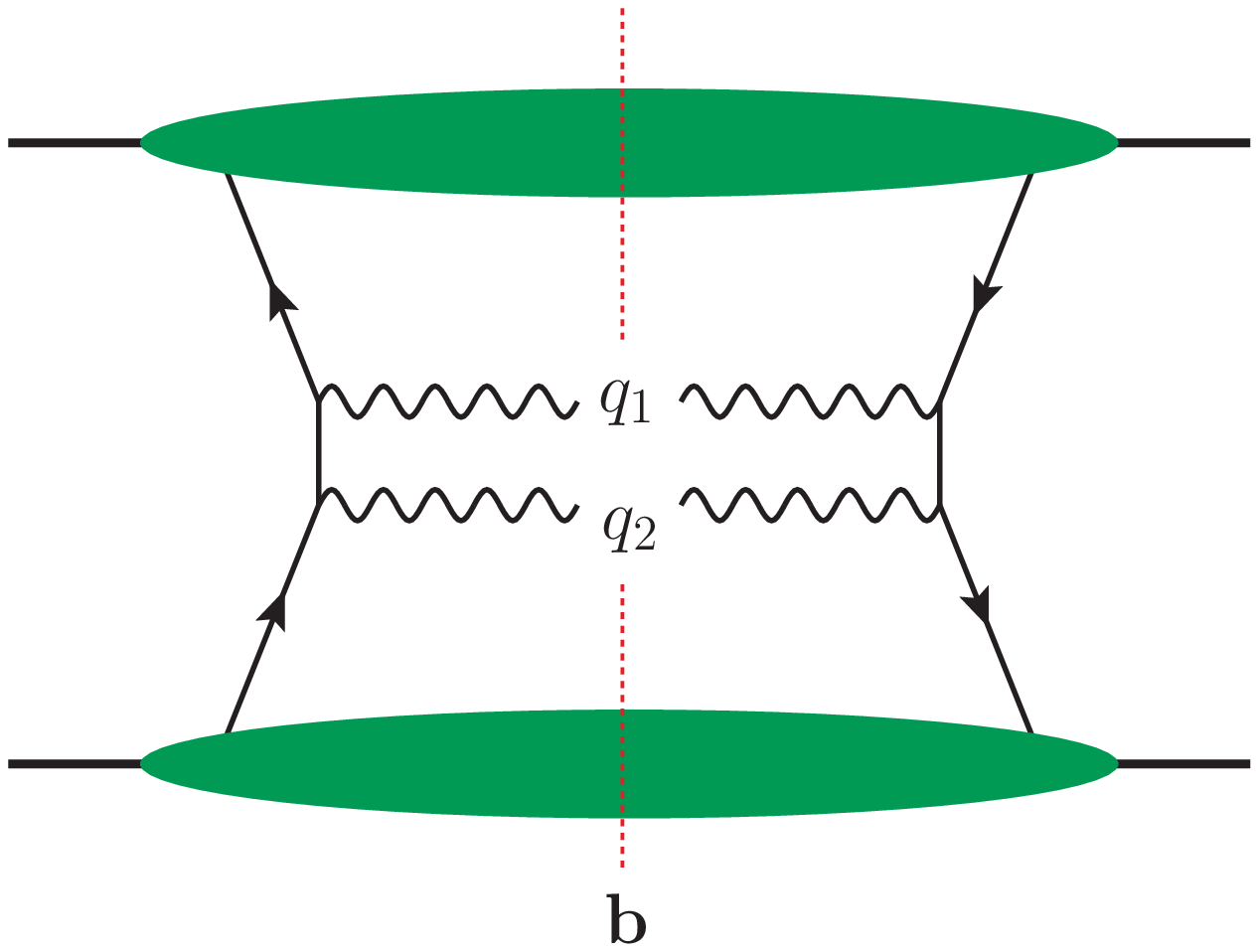}
\hspace{1em}
\end{center}
\caption{\label{fig:scatter} Graphs for the production of two gauge bosons
  by double (a) or single (b) hard scattering.  The dotted line denotes
  the final-state cut.  The decays of the gauge bosons into
  fermion-antifermion pairs are not shown for simplicity.}
\end{figure*}

We begin by sketching the derivation of the cross section formula for
double parton scattering at tree level.  For definiteness we take two
colliding protons and consider the case where the two partons coming from
one of the protons are quarks.  The corresponding graph is shown in
Fig.~\ref{fig:scatter}a, which also specifies our assignment of momentum
variables.

We use light-cone coordinates $v^\pm = (v^0 \pm v^3)/\sqrt{2}$ and
$\tvec{v} = (v^1, v^2)$ for any four-vector $v$ and choose a reference
frame where $p$ moves fast to the right and $\bar{p}$ fast to the left,
with transverse momenta $\tvec{p} = \bar{\tvec{p}} = \tvec{0}$.  We
consider kinematics where the invariant masses of the bosons are large and
where their transverse momenta are much smaller, i.e.\ we require $q_T \ll
Q$ with $\tvec{q}_1^2 \sim \tvec{q}_2^2 \sim q_T^2$ and $q_1^2 \sim q_2^2
\sim Q^2$.

The lower blob in Fig.~\ref{fig:scatter}a is described by the correlation
function for two quarks in a proton,
\begin{align}
  \label{Phi-quarks}
& \Phi_{\alpha_1 \beta_1 \alpha_2 \beta_2}(k_1, k_2, r)
 = \int \frac{d^4 z_1}{(2\pi)^4}\, \frac{d^4 z_2}{(2\pi)^4}\,
        \frac{d^4 y}{(2\pi)^4}
\nonumber \\
& \quad \times e^{i z_1 k_1 + i z_2 k_2 - i y r}\,
    \big\langle p \big| 
    \bar{T} \Bigl[\, \bar{q}_{\beta_1}\bigl(y - \half z_1\bigr)\ms
                     \bar{q}_{\beta_2} \bigl(- \half z_2\bigr) \Bigr]
\nonumber \\
& \qquad \times
    T \Bigl[\, q_{\alpha_2} \bigl(\half z_2\bigr)\ms
               q_{\alpha_1}\bigl(y + \half z_1\bigr) \Bigr]
    \big| p \big\rangle \,.
\end{align}
Here $T$ denotes time ordering and $\bar{T}$ anti-time ordering, as
appropriate for fields in the scattering amplitude or its conjugate.  The
association between momenta and field positions becomes clear if one
rewrites $z_1 k_1 + z_2 k_2 - y r = (y + \half z_1) (k_1 - \half r) +
\half z_2 (k_2 + \half r) + \half z_2 (k_2 - \half r) - (y - \half z_1)
(k_1 + \half r)$ in the Fourier exponent.
For now we gloss over the flavor and color structure of $\Phi$, which will
be discussed in Sections~\ref{sec:interference} and ~\ref{sec:color}.
Throughout this work we consider unpolarized protons, so that an average
over the proton spin in \eqref{Phi-quarks} is understood.  The correlation
function $\bar{\Phi}$ for two antiquarks in a proton is defined in analogy
to \eqref{Phi-quarks}, with interchanged roles of the $q$ and $\bar{q}$
fields.  The contribution of graph \ref{fig:scatter}a to the cross section
then reads
\begin{align}
  \label{Xsect-start}
& \frac{d\sigma \ms|_{\text{\protect\ref{fig:scatter}a}}}{
        d^4 q_1\, d^4 q_2}
= \frac{1}{S}\, \frac{1}{4p\bar{p}}\,
  \biggl[\, \prod_{i=1}^{2}
     \int \! d^4k_i\, d^4\bar{k}_i\;
          \delta^{(4)}(q_i - k_i - \bar{k}_i) \biggr]
\nonumber \\
 &\quad \times
  (2\pi)^4 \int d^4r\, d^4\bar{r}\; \delta^{(4)}(r + \bar{r}) \,
  \biggl[\, \prod_{i=1}^{2}
           H_{i,\, \beta_i \alpha_i \bar\beta_i \bar\alpha_i} \biggr]
\nonumber \\
 &\quad \times
   \Phi_{\alpha_1 \beta_1 \alpha_2 \beta_2}(k_1, k_2, r)\,
   \bar{\Phi}_{\bar\alpha_1 \bar\beta_1 \bar\alpha_2 \bar\beta_2}
             (\bar{k}_1, \bar{k}_2, \bar{r}) \,,
  \phantom{\biggl[ \biggr]} \hspace{-2em}
\end{align}
where $H_i (k_i,\bar{k}_i, r,\bar{r})$ is the squared amplitude for each
of the two hard-scattering processes.  Repeated Dirac indices $\alpha_1$,
$\beta_1$, etc.\ are to be summed over.  The statistical factor $S$ is 2
if the final states of the two hard scatters are identical and 1
otherwise.

To proceed, we make the same approximations as in processes with a single
hard scattering:
\begin{enumerate}
\item use that the minus-momenta of right-moving partons and the
  plus-momenta of left-moving partons are small compared with the large
  scale $Q$.  The constraint $\delta^{(4)}(r + \bar{r})$ forces $r^+$ and
  $\bar{r}^{\ms-}$ to be small, although by general scaling arguments they
  could both be large, and the constraint $\delta^{(4)}(q_i - k_i -
  \bar{k}_i)$ leads to $k_i^+ \approx q_i^+$ and $\bar{k}_i^- \approx
  q_i^-$.  Defining
  \begin{align}
    x_i^{} &= q_i^+/p^+ \,, &
    \bar{x}_i^{} &= q_i^-/\bar{p}^{\ms-} \,,
  \end{align}
  we thus find that the parton momentum fractions $(k_i^+ \pm \half r^+)
  /p^+ \approx x_i$ and $(\bar{k}_i^- \pm \half \bar{r}^-) /\bar{p}^-
  \approx \bar{x}_i$ are fixed by the final-state kinematics.  Note that
  this does not hold for the transverse parton momenta, since $\tvec{q}_i$
  receives contributions from both $\tvec{k}_i \pm \half\tvec{r}$ and
  $\bar{\tvec{k}}_i \pm \half\bar{\tvec{r}}$.
\item neglect small plus- and minus-components, as well as all transverse
  momenta in the squared parton-level amplitudes $H_i$, which then depend
  only on $q_i^+$ and $q_i^-$, i.e.\ on quantities of order $Q$.  After
  this approximation, the integrations over $k_i^-$ and $r^-$ in
  \eqref{Xsect-start} only concern the factor $\Phi$, and those over
  $\bar{k}_i^+$ and $\bar{r}^{\ms+}$ only concern the factor $\bar\Phi$.
\item perform a Fierz transformation for the index pairs $(\alpha_i,
  \beta_i)$ and $(\bar\alpha_i, \bar\beta_i)$ in \eqref{Xsect-start}.  The
  correlation function $\Phi$ is then multiplied with Dirac matrices that
  can carry Lorentz indices.  One has to retain only those terms with the
  maximum number of plus indices, since the momenta on which $\Phi$
  depends are largest in the plus-direction.  For $\bar\Phi$ the
  dominating terms have the maximum number of minus indices.
\end{enumerate}
After these steps we find that the two-quark distributions required to
describe the graph \ref{fig:scatter}a read
\begin{align}
  \label{F-def}
& F_{a_1,a_2}(x_1, x_2, \tvec{k}_1, \tvec{k}_2, \tvec{r})
= (2\pi)^{3}\, 2p^+ \!\! \int dk_1^-\, dk_2^-\, dr^-
\nonumber \\
 & \quad \times
  (\Gamma_{a_1})_{\beta_1\alpha_1}\,
  (\Gamma_{a_2})_{\beta_2\alpha_2} 
    \Phi_{\alpha_1 \beta_1 \alpha_2 \beta_2}(k_1, k_2, r)
  \,\bigg|_{\substack{k_i^+ = \, x_i^{} p^+ \\ r^+ = \, 0 \hfill}}
\nonumber \\
 &= 2p^+ \biggl[\, \prod_{i=1}^2
       \int \frac{dz_i^- d^2\tvec{z}_i^{}}{(2\pi)^3}\,
       e^{i (x_i^{} z_i^- p^+ - \tvec{z}_i^{} \tvec{k}_i^{})}
    \biggr]\,
\nonumber \\
 &\quad \times \int dy^- d^2\tvec{y}\; e^{i \tvec{y} \tvec{r}}\,
    \big\langle p \big|\,
    \mathcal{O}_{a_2}(0, z_2)\ms \mathcal{O}_{a_1}(y, z_1)
    \big| p \big\rangle
\end{align}
with bilinear operators
\begin{align}
  \label{quark-bilinears}
\mathcal{O}_{a}(y, z) &= \bar{q}(y - \half z)\,
  \Gamma_a \, q(y + \half z) \Big|_{z^+ = y^+ = 0} \,.
\end{align}
Here $a = q, \Delta q, \delta q$ labels the quark polarization and
\begin{align}
  \label{Gamma-twist-2}
\Gamma_{q}          &= \half \gamma^+ \,, &
\Gamma_{\Delta q}   &= \half \gamma^+ \gamma_5 \,, &
\Gamma_{\delta q}^j &= \half i \sigma^{j +} \gamma_5
\end{align}
with $j=1,2$.  The operators in \eqref{quark-bilinears} are well-known
from the definitions of single-parton densities for unpolarized,
longitudinally polarized and transversely polarized quarks, see e.g.\
\cite{Ralston:1979ys,Tangerman:1994eh}.  Analogous definitions hold for
antiquarks and for a left-moving hadron.  The cross section in
\eqref{Xsect-start} can finally be written as
\begin{align}
  \label{X-sect-mom}
& \frac{d\sigma \ms|_{\text{\protect\ref{fig:scatter}a}}}{
        \prod_{i=1}^2 dx_i\, d\bar{x}_i\, d^2\tvec{q}{}_i}
= \frac{1}{S}\, 
  \sum_{\substack{a_1, a_2 = q, \Delta q, \delta q \\[0.1ex]
          \bar{a}_1, \bar{a}_2 = \bar{q}, \Delta\bar{q}, \delta\bar{q}}}
  \biggl[\, \prod_{i=1}^{2} \hat{\sigma}_{i, a_i \bar{a}_i}(q_i^2)\,
\nonumber \\
& \quad \times  
  \int d^2\tvec{k}_i\, d^2\bar{\tvec{k}}_i\;
    \delta^{(2)}(\tvec{q}{}_i - \tvec{k}_i - \bar{\tvec{k}}_i) \biggr]
\nonumber \\[0.2em]
& \quad \times \int \frac{d^2\tvec{r}}{(2\pi)^2}\;
  F_{a_1, a_2}(x_i, \tvec{k}_i, \tvec{r})\, 
  F_{\bar{a}_1, \bar{a}_2}(\bar{x}_i,
                           \bar{\tvec{k}}_i, -\tvec{r}) \,,
  \phantom{\int} \hspace{-1em}
\end{align}
where here and in the following we write $F(x_i, \tvec{k}_i, \tvec{r})$
instead of $F(x_1, x_2, \tvec{k}_1, \tvec{k}_2, \tvec{r})$ for brevity.
The parton-level cross sections are given by
\begin{align}
  \label{quark-X-sect}
\hat{\sigma}_{i, a \bar{a}} &= \frac{1}{2 q_i^2}\,
  \bigl[ P_{a}(k_i) \bigr]{}_{\alpha\beta}\,
  \bigl[ P_{\bar{a}}(\bar{k}_i) \bigr]{}_{\bar{\alpha} \bar{\beta}}\;
  H_{i, \beta\alpha \bar{\beta} \bar{\alpha}}
\end{align}
with quark spin projectors $P_{q}(k) = \smash{\half} k^+ \gamma^-$,
$P_{\Delta q}(k) = \smash{\half} \gamma_5\ms k^+ \gamma^-$,
$\smash{P^j_{\delta q}(k) = \half} \gamma_5\ms k^+ \gamma^- \gamma^j_{}$
and corresponding antiquark spin projectors $P_{\bar{a}}$.  It is
understood that for each $a_i=\delta q$ both $F_{a_1, a_2}$ and
$\hat{\sigma}_{i, a \bar{a}}$ carry extra indices $j$ associated with the
direction of the transverse quark polarization.  Corresponding remarks
hold for $\bar{a}_i=\delta \bar{q}$.

The difference $\tvec{r}$ of transverse parton momenta can be replaced by
the Fourier conjugate position $\tvec{y}$, both in the distributions
\begin{align}
  \label{F-mixed}
& F_{a_1,a_2}(x_i, \tvec{k}_i, \tvec{y})
\nonumber \\
 &\quad = \int \frac{d^2\tvec{r}}{(2\pi)^2}\, e^{-i \tvec{y} \tvec{r}}
  F_{a_1,a_2}(x_i, \tvec{k}_i, \tvec{r})
\nonumber \\
 &\quad= \biggl[\, \prod_{i=1}^2
       \int \frac{dz_i^- d^2\tvec{z}_i^{}}{(2\pi)^3}\,
       e^{i (x_i^{} z_i^- p^+ - \tvec{z}_i^{} \tvec{k}_i^{})}
    \biggr]
\nonumber \\
 &\qquad \times 2 p^+\!\! \int dy^-\,
    \big\langle p \big|\,
    \mathcal{O}_{a_2}(0, z_2)\ms \mathcal{O}_{a_1}(y, z_1) \big| p
    \big\rangle
\end{align}
and in the cross section
\begin{align}
  \label{X-sect-mixed}
& \frac{d\sigma \ms|_{\text{\protect\ref{fig:scatter}a}}}{
        \prod_{i=1}^2 dx_i\, d\bar{x}_i\, d^2\tvec{q}{}_i}
= \frac{1}{S}\, 
  \sum_{\substack{a_1, a_2 = q, \Delta q, \delta q \\[0.1ex]
          \bar{a}_1, \bar{a}_2 = \bar{q}, \Delta\bar{q}, \delta\bar{q}}}
  \biggl[\, \prod_{i=1}^{2} \hat{\sigma}_{i, a_i \bar{a}_i}(q_i^2)\,
\nonumber \\
& \quad \times  
  \int d^2\tvec{k}_i\, d^2\bar{\tvec{k}}_i\;
    \delta^{(2)}(\tvec{q}{}_i - \tvec{k}_i - \bar{\tvec{k}}_i) \biggr]
\nonumber \\[0.2em]
& \quad \times \int d^2\tvec{y}\,
  F_{a_1, a_2}(x_i, \tvec{k}_i, \tvec{y})\;
  F_{\bar{a}_1, \bar{a}_2}(\bar{x}_i, \bar{\tvec{k}}_i, \tvec{y}) \,.
  \phantom{\int}
\end{align}
Integration of the cross section over $\tvec{q}_1$ and $\tvec{q}_2$ leads
to collinear (i.e.\ transverse-momentum integrated) two-parton densities
\begin{align}
  \label{F-coll}
& F_{a_1,a_2}(x_i, \tvec{y})
 = \int d^2\tvec{k}_1\, d^2\tvec{k}_2\, 
   F_{a_1,a_2}(x_i, \tvec{k}_i, \tvec{y}) \,.
\end{align}
The corresponding cross section formula is the basis for the phenomenology
of multiple interactions and has been used for a long time.  It was
derived in \cite{Paver:1982yp} for scalar partons and in
\cite{Mekhfi:1983az} for quarks, in ways similar to the one we just
sketched.

As we will discuss in Section~\ref{sec:evolution} the integral in
\eqref{F-coll} diverges logarithmically at order $\alpha_s$ and requires
appropriate regularization.  Up to this caveat, which also applies to
single-parton densities, $F(x_i, \tvec{y})$ gives the probability for
finding two partons with momentum fractions $x_1$ and $x_2$ at relative
transverse distance $\tvec{y}$ in a proton.  By contrast, $F(x_i,
\tvec{k}_i, \tvec{y})$ depends on both transverse-momentum and
transverse-position arguments for the quarks and does not admit a
probability interpretation due to the uncertainty principle.  Instead,
$F(x_i, \tvec{k}_i, \tvec{y})$ has the structure of a Wigner distribution
\cite{Hillery:1984} for the transverse degrees of freedom: the integral
\eqref{F-coll} gives the probability to find two partons at transverse
distance $\tvec{y}$, whereas the integral $\int d^2\tvec{y}\, F(x_i,
\tvec{k}_i, \tvec{y}) = F(x_i, \tvec{k}_i, \tvec{r}=\tvec{0})$ gives the
probability to find two partons with transverse momenta $\tvec{k}_1$ and
$\tvec{k}_2$.  A similar discussion for generalized parton distributions
can be found in \cite{Belitsky:2003nz}.  Using \eqref{quark-bilinears} and
\eqref{F-mixed} we can identify $\tvec{k}_1$ and $\tvec{k}_2$ as the
``average'' transverse momenta of the two partons and $\tvec{y}$ as their
``average'' transverse distance, where the ``average'' refers to the
scattering amplitude and its complex conjugate (see
Fig.~\ref{fig:scatter}a).  In this sense the cross section formula
\eqref{X-sect-mixed} describes two hard-scattering processes where a quark
and an antiquark with average transverse momenta $\tvec{k}_i$ and
$\bar{\tvec{k}}_i$ annihilate into a gauge boson with transverse momentum
$\tvec{q}{}_i = \tvec{k}_i + \bar{\tvec{k}}_i$.  The two annihilation
processes occur at an average transverse distance $\tvec{y}$, which equals
the average transverse distance of the two quarks in one proton and of the
two antiquarks in the other.  If the cross section is integrated over
$\tvec{q}_1$ and $\tvec{q}_2$, the specification of ``average'' for the
distance $\tvec{y}$ can be dropped.

It is gratifying to find such an intuitive physical interpretation, which
we will further develop in Section~\ref{sec:impact}.  On the other hand,
we emphasize that \eqref{X-sect-mixed} and its $\tvec{q}_i$-integrated
version can be derived from Feynman graphs using standard hard-scattering
approximations, without any need to appeal to classical or semi-classical
arguments.

One can also Fourier transform the two-parton distributions w.r.t.\ the
transverse momenta $\tvec{k}_i$,
\begin{align}
  \label{F-position}
& F_{a_1,a_2}(x_i, \tvec{z}_i, \tvec{y})
\nonumber \\
 &\quad = \biggl[\, \prod_{i=1}^2
      \int d^2\tvec{k}_i\, e^{i \tvec{z}_i \tvec{k}_i}
    \biggr]  F_{a_1,a_2}(x_i, \tvec{k}_i, \tvec{y})
\nonumber \\
 &\quad =  2 p^+\!\! \int dy^-\, \frac{dz_1^-}{2\pi}\,
       \frac{dz_2^-}{2\pi}\,
       e^{i (x_1^{} z_1^- + x_2^{} z_2^-) p^+}
\nonumber \\
 &\qquad \times
    \big\langle p \big|\,
    \mathcal{O}_{a_2}(0, z_2)\ms \mathcal{O}_{a_1}(y, z_1) \big| p
    \big\rangle \,.
  \phantom{\frac{1}{1}}
\end{align}
This form is most suitable for the resummation of Sudakov logarithms, as
is well-known for single-parton distributions \cite{Collins:1981uk}.
Indeed, the cross section formula
\begin{align}
  \label{X-sect-position}
& \frac{d\sigma \ms|_{\text{\protect\ref{fig:scatter}a}}}{
        \prod_{i=1}^2 dx_i\, d\bar{x}_i\, d^2\tvec{q}{}_i}
 = \frac{1}{S}
\nonumber \\
& \quad \times
  \sum_{\substack{a_1, a_2 = q, \Delta q, \delta q \\[0.1ex]
          \bar{a}_1, \bar{a}_2 = \bar{q}, \Delta\bar{q}, \delta\bar{q}}}
  \biggl[\, \prod_{i=1}^{2} \hat{\sigma}_{i, a_i \bar{a}_i}(q_i^2)\,
    \int \frac{d^2\tvec{z}_i}{(2\pi)^2}\, e^{-i \tvec{z}_i \tvec{q}{}_i}
  \biggr]
\nonumber \\[0.1em]
& \quad \times \int d^2\tvec{y}\,
  F_{a_1, a_2}(x_i, \tvec{z}_i, \tvec{y})\;
  F_{\bar{a}_1,\bar{a}_2}(\bar{x}_i, \tvec{z}_i, \tvec{y})
  \phantom{\int} \hspace{-1em}
\end{align}
is reminiscent of the one for single Drell-Yan production
\cite{Collins:1984kg}.
Up to terms of order $\alpha_s$ related to the regularization of
logarithmic divergences, the collinear distributions $F(x_i, \tvec{y})$
are obtained from $F(x_i, \tvec{z}_i, \tvec{y})$ by setting $\tvec{z}_i =
\tvec{0}$.  We note that by taking the complex conjugate of
\eqref{F-mixed} one easily finds that $F(x_i, \tvec{k}_i, \tvec{y})$ is
real valued, whereas $F(x_i, \tvec{k}_i, \tvec{r})$ and $F(x_i,
\tvec{z}_i, \tvec{y})$ can have complex phases.

The graph in Fig.~\ref{fig:scatter}a involves a two-quark distribution
$F_{a_1, a_2}$ in proton $p$.  The cross section receives further
contributions where the two partons in proton $p$ are both antiquarks, or
where one is a quark and the other an antiquark.  The definitions of
quark-antiquark distributions $F_{a_1, \bar{a}_2}$ and $F_{\bar{a}_1,
  a_2}$ and the associated cross section formulae are close analogs of the
expressions given above.  There are however further contributions, which
will be discussed in Section~\ref{sec:interference}.

The preceding results can be generalized to hard-scattering processes
initiated by gluons, such as gluon-gluon fusion into a Higgs boson via a
top-quark loop.  Two-gluon distributions are defined as in
\eqref{F-mixed}, but with an extra factor of $1/(x_1 p^+\ms x_2\ms p^+)$
on the r.h.s.\ and with operators that are bilinear in the gluon field
strength $G^{\mu\nu}$,
\begin{align}
  \label{gluon-bilinears}
\mathcal{O}_{a}^{}(y, z) &=
\Pi_{a}^{jj'}  G^{+ j'}(y - \half z)\, G^{+ j}(y + \half z) \,,
\end{align}
where $a = g, \Delta g, \delta g$ are polarization labels and
\begin{align}
\Pi_{g}^{jj'} &= \delta^{jj'} \, , 
\qquad\qquad
\Pi_{\Delta g}^{jj'} = i\epsilon^{jj'} \, ,
\nonumber \\[0.1em]
\bigl[ \Pi_{\delta g}^{\ms l\ms l'} \bigr]{}^{jj'} &=
\half \bigl(
  \delta^{jl} \delta^{j'l'} + \delta^{jl'} \delta^{j'l}
- \delta^{jj'} \delta^{ll'} \bigr) \,.
\end{align}
The operators $\mathcal{O}_{g}$ and $\mathcal{O}_{\Delta g}$ appear in the
usual densities for unpolarized and longitudinally polarized gluons.
$\mathcal{O}_{\delta g}^{\ms l\ms l'}$ describes linear gluon polarization
(or equivalently gluon helicity flip by two units) and has been discussed
in \cite{Jaffe:1989xy,Belitsky:2000jk}.


\section{Power behavior}
\label{sec:power}

A pair of electroweak gauge bosons can be produced by two hard scatters,
but also by a single one.  An example graph is shown in figure
\ref{fig:scatter}b, and the corresponding cross section formula reads
\begin{align}
  \label{single-hard-diff}
& \frac{d\sigma \ms|_{\text{\protect\ref{fig:scatter}b}}}{
        \prod_{i=1}^2 dx_i\, d\bar{x}_i\, d^2\tvec{q}{}_i}
 = \frac{d\hat{\sigma}}{dx_1\, d\bar{x}_1\, d^2\tvec{q}{}_1}
   \int d^2\tvec{k}\, d^2\bar{\tvec{k}}\;
\nonumber \\
& \quad\times
    \delta^{(2)}(\tvec{q}_1 + \tvec{q}_2 - \tvec{k} - \bar{\tvec{k}})\,
    f_q(x,\tvec{k})\, f_{\bar{q}}(\bar{x}, \bar{\tvec{k}}) \,,
  \phantom{\frac{1}{1}} \hspace{-2em}
\end{align}
where $x = x_1+x_2$, $\bar{x} = \bar{x}_1+\bar{x}_2$ and $\hat{\sigma}$ is
the cross section for $q\bar{q}$ annihilation into two gauge bosons.

Dimensional analysis of \eqref{X-sect-mom} and \eqref{single-hard-diff}
reveals that
\begin{align}
  \label{pow-1}
\frac{d\sigma}{\prod_{i=1}^2 dx_i\, d\bar{x}_i\, d^2\tvec{q}{}_i}
 \sim \frac{1}{Q^4 \Lambda^2}
\end{align}
for both the single and double hard-scattering mechanisms.  Here the small
scale $\Lambda^2$ represents $q_T^2$ or the scale of non-perturbative
interactions, whichever is larger.  To obtain \eqref{pow-1} one uses that
parton distributions do not depend on the hard scale $Q^2$ (except for
logarithms, which are discarded when counting powers), whereas
hard-scattering cross sections are independent of $\Lambda^2$.  We thus
obtain an important result: multiple hard scattering is \emph{not} power
suppressed in cross sections that are sufficiently differential in
transverse momenta.

The situation changes when one integrates over $\tvec{q}_1$ and
$\tvec{q}_2$.  In the double-scattering mechanism both transverse boson
momenta are restricted to order $\Lambda$ since they result from the
transverse momenta of the annihilating partons.  For a single hard
scattering one has $|\tvec{q}_1 + \tvec{q}_2| \sim \Lambda$, whereas the
individual transverse boson momenta can be as large as $Q$.  Due to this
phase space effect one obtains
\begin{align}
\frac{d\sigma \ms|_{\text{\protect\ref{fig:scatter}a}}}{
      \prod_{i=1}^2 dx_i\, d\bar{x}_i}
& \sim\, \frac{\Lambda^2}{Q^4} \, , & 
\frac{d\sigma \ms|_{\text{\protect\ref{fig:scatter}b}}}{
      \prod_{i=1}^2 dx_i\, d\bar{x}_i}
& \sim\, \frac{1}{Q^2} \,.
\end{align}
In the transverse-momentum integrated cross section multiple hard
scattering is thus power suppressed.  This is in fact required for the
validity of the usual collinear factorization formulae, which describe
only the single-hard-scattering contribution.

The power behavior just discussed remains the same for parton-level
processes initiated by gluons instead of quarks, and for final states
other than gauge bosons.


\section{Impact parameter}
\label{sec:impact}

\begin{figure}
\begin{center}
\includegraphics[width=0.44\textwidth]{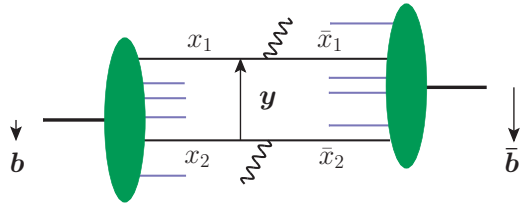}
\end{center}
\caption{\label{fig:impact-scatter} Visualization of the cross section
  formula \protect\eqref{X-sect-impact} when $\tvec{q}_1$ and $\tvec{q}_2$
  are integrated over.}
\end{figure}

The distributions $F(x_i, \tvec{z}_i, \tvec{y})$ depend on spatial
transverse coordinates for the quarks but still refer to a proton with
definite (zero) transverse momentum.  A representation purely in impact
parameter space can be obtained using the methods of
\cite{Soper:1976jc,Burkardt:2002hr,Diehl:2002he}, where impact parameter
densities for a single parton are constructed from generalized parton
distributions.  To this end we first define non-forward distributions
$F(x_i, \tvec{z}_i, \tvec{y}; \tvec{\Delta})$ as in \eqref{F-position} but
between proton states $\langle p^+, \half\tvec{\Delta}|$ and $|p^+,
-\half\tvec{\Delta}\rangle$ with different transverse momenta.
Introducing the wave packet
\begin{align}
  \label{impact-states}
|p^+, \tvec{b} \rangle = \int \frac{d^2\tvec{p}}{(2\pi)^2}\,
   e^{-i \tvec{b} \tvec{p}}\, |p^+, \tvec{p} \rangle \,,
\end{align}
which describes a proton with definite transverse position $\tvec{b}$, one
can show that
\begin{align}
  \label{F-impact}
& 2 p^+\!\! \int dy^-\, \frac{dz_1^-}{2\pi}\, \frac{dz_2^-}{2\pi}\,
       e^{i (x_1^{} z_1^- + x_2^{} z_2^-) p^+}\,
  \nonumber \\
&\quad \times \big\langle p^+, -\tvec{b} -\half\tvec{d}
    \big|\, \mathcal{O}_{a_2}(0, z_2)\ms \mathcal{O}_{a_1}(y, z_1)
    \big| p^+, -\tvec{b} +\half\tvec{d}
    \big\rangle
  \phantom{\frac{1}{1}}
\nonumber \\
 &= \delta^{(2)}( \tvec{d} - x_1 \tvec{z}_1 - x_2 \tvec{z}_2)\,
    F_{a_1,a_2}\bigl(x_i, \tvec{z}_i, \tvec{y}; \tvec{b} \big)
    \phantom{\int}
\end{align}
with
\begin{align}
& F_{a_1,a_2}\bigl(x_i, \tvec{z}_i, \tvec{y}; \tvec{b} \big)
\nonumber \\
&\quad
= \int \frac{d^2\tvec{\Delta}}{(2\pi)^2}\; e^{-i \tvec{b} \tvec{\Delta}}
  F_{a_1,a_2}\bigl(x_i, \tvec{z}_i, \tvec{y}; \tvec{\Delta} \big) \,.
\end{align}
The difference $\tvec{d}$ in the transverse positions of the two proton
states is a consequence of Lorentz invariance, see~\cite{Diehl:2002he}.
Integrating $F(x_i, \tvec{z}_i, \tvec{y}; \tvec{b})$ over $\tvec{b}$ one
recovers the distributions $F(x_i, \tvec{z}_i, \tvec{y})$, so that the
cross section \eqref{X-sect-position} can be cast into the form
\begin{align}
  \label{X-sect-impact}
& \frac{d\sigma \ms|_{\text{\protect\ref{fig:scatter}a}}}{
        \prod_{i=1}^2 dx_i\, d\bar{x}_i\, d^2\tvec{q}{}_i}
 = \frac{1}{S}\,
  \sum_{\substack{a_1, a_2 = q, \Delta q, \delta q \\[0.1ex]
          \bar{a}_1, \bar{a}_2 = \bar{q}, \Delta\bar{q}, \delta\bar{q}}}
  \biggl[\, \prod_{i=1}^{2} \hat{\sigma}_{i, a_i \bar{a}_i}(q_i^2)
\nonumber \\
& \quad \times
    \int \frac{d^2\tvec{z}_i}{(2\pi)^2}\, e^{-i \tvec{z}_i \tvec{q}_i}
  \biggr]  \int d^2\tvec{y}\; d^2\tvec{b}\; d^2\bar{\tvec{b}}
\nonumber \\
& \quad \times
  F_{a_1, a_2}(x_i, \tvec{z}_i, \tvec{y}; \tvec{b})\;
  F_{\bar{a}_1,\bar{a}_2}(\bar{x}_i, \tvec{z}_i, \tvec{y};
     \bar{\tvec{b}}) \,,
  \phantom{\int} \hspace{-1em}
\end{align}
which has a simple geometric interpretation in impact parameter space.
Taking the average of transverse positions in the amplitude and its
conjugate as in Section~\ref{sec:tree}, one identifies $\tvec{y}$ as the
average distance between the two scattering partons, $\tvec{b}$ as the
average distance between parton 2 and the right-moving proton, and
$\bar{\tvec{b}}$ as the average distance between parton 2 and the
left-moving proton.  This is illustrated in Fig.~\ref{fig:impact-scatter}
for the case where the cross section is integrated over $\tvec{q}_i$, so
that $\tvec{z}_i = \tvec{0}$ and the positions in the amplitude and its
conjugate coincide.  

The $\tvec{q}{}_i$-integrated version of \eqref{X-sect-impact} has
previously been derived in \cite{Calucci:2009ea}.  Given the work in
\cite{Paver:1982yp,Mekhfi:1983az} and \cite{Calucci:2009ea}, we disagree
with the statement in \cite{Blok:2010ge} that the impact parameter picture
of multiparton interactions has until now been based on ``semi-intuitive
reasoning''.

\section{Interference contributions}
\label{sec:interference}

\begin{figure}
\begin{center}
\includegraphics[width=0.49\textwidth]{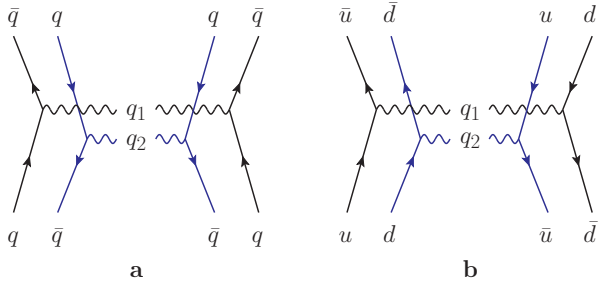}
\end{center}
\caption{\label{fig:interference} Example graphs for interference terms in
  fermion number (a) and in quark flavor (b).  The blobs indicating
  two-parton distributions are not shown for simplicity.  Labels $q$ and
  $\bar{q}$ indicate whether a line is represented by a quark field or a
  conjugate quark field in the corresponding matrix element.  Momenta are
  to be assigned as in Fig.~\protect\ref{fig:scatter}a.}
\end{figure}

The multiparton distributions discussed so far have an interpretation as
probabilities or as pseudo-probabilities in the sense of Wigner
distributions.  However, there are contributions to the cross section
which involve distributions that are interference terms rather than
probabilities.  Figure~\ref{fig:interference}a shows an example where the
parton with momentum fraction $x_1$ is a quark in the scattering amplitude
and an antiquark in the conjugate scattering amplitude.  Such interference
terms in the fermion number of the partons have no equivalent in single
hard-scattering processes, where they are forbidden by fermion number
conservation.  Their contribution to the cross section has the same
structure and in particular the same power behavior as the contributions
discussed in Section~\ref{sec:tree}.  To the best of our knowledge such
interference terms are not included in existing phenomenology.

Let us introduce a shorthand notation
\begin{align}
  \label{shorthand}
& \mat{\varphi_4\, \varphi_3\, \varphi_2\, \varphi_1}
 = \biggl[\, \prod_{i=1}^2
       \int \frac{dz_i^-\, d^2\tvec{z}_i^{}}{(2\pi)^3}\,
       e^{i (x_i^{} z_i^- p^+ - \tvec{z}_i^{} \tvec{k}_i^{})}
    \biggr]\,
\nonumber \\
 &\quad \times
  2p^+ \!\! \int dy^-
    \big\langle p \big|\,
       \varphi(y - \half z_1)\, \varphi(- \half z_2)\,
\nonumber \\
 &\qquad\qquad \times
       \varphi(\half z_2)\, \varphi(y + \half z_1)
    \big| p \big\rangle \Big|_{z_1^+ = z_2^+ = y^+_{\phantom{1}} = 0}
\end{align}
for the Fourier transformed matrix element of a product of field operators
$\varphi$, with indices assigned as
\begin{alignat*}{3}
1 &\;\leftrightarrow\; y + \half z_1 
 &&\;\leftrightarrow\; \text{mom.\ fract.\ $x_1$ in amplitude}
\nonumber \\
2 &\;\leftrightarrow\; \hspace{1.8em} \half z_2
 &&\;\leftrightarrow\; \text{mom.\ fract.\ $x_2$ in amplitude}
\nonumber \\
3 &\;\leftrightarrow\; \phantom{y} - \half z_2
 &&\;\leftrightarrow\; \text{mom.\ fract.\ $x_2$ in conjugate
                             ampl.}
\nonumber \\
4 &\;\leftrightarrow\; y - \half z_1
 &&\;\leftrightarrow\; \text{mom.\ fract.\ $x_1$ in conjugate
                             ampl.}
\end{alignat*}
The numbering corresponds to the order of the parton lines in
figures~\ref{fig:scatter}a and \ref{fig:interference}, from left to right.
We then respectively have
\begin{align}
F_{a_1, a_2}(x_i, \tvec{k}_i, \tvec{y}) &=
  \mat{(\bar{q}_3\ms \Gamma_{a_2}\ms q_2)\,
       (\bar{q}_4\ms \Gamma_{a_1}\ms q_1)} \,,
\nonumber \\
F_{a_1, \bar{a}_2}(x_i, \tvec{k}_i, \tvec{y}) &=
  \mat{(\bar{q}_2\ms \Gamma_{\bar{a}_2}\ms q_3)\,
       (\bar{q}_4\ms \Gamma_{a_1}\ms q_1)}
\end{align}
for a two-quark and a quark-antiquark distribution, whereas the
quark-antiquark interference distribution associated with the lower part
of figure \ref{fig:interference}a is given by
\begin{align}
  \label{interf-dist}
I_{a_1,\bar{a}_2}(x_i, \tvec{k}_i, \tvec{y}) &=
  \mat{(\bar{q}_2\ms \Gamma_{\bar{a}_2}\ms q_4)\,
       (\bar{q}_3\ms \Gamma_{a_1}\ms q_1)} \,.
\end{align}

So far we have not paid attention to the quark flavor structure of the
double-scattering process.  One readily finds that there are also
interference terms in the flavor quantum numbers of the partons.  An
example is given in Fig.~\ref{fig:interference}b, where the parton with
momentum fraction $x_1$ is a $u$ quark in the amplitude and a $d$ quark in
the conjugate amplitude.  The corresponding distribution is given by the
matrix element $\mat{(\bar{u}_3\ms \Gamma_{a_2}\ms d_2)\, (\bar{d}_4\ms
  \Gamma_{a_1}\ms u_1)}$.


\section{Color structure}
\label{sec:color}

In contrast to single-parton densities, where two parton fields are always
coupled to a color singlet, multiparton distributions have a nontrivial
color structure, which we now discuss.  A color decomposition of two-quark
distributions is given by
\begin{align}
  \label{quark-color-decomp}
& F_{jj', kk'} =
  \mat{( \bar{q}_{3,k'}\ms \Gamma_{a_2}\ms q_{2,k} )\,
       ( \bar{q}_{4,j'}\ms \Gamma_{a_1}\ms q_{1,j} )}
\nonumber \\
&\quad = \frac{1}{N^2} \Bigl[ \sing{F}\, \delta_{jj'} \delta_{kk'}
     + \frac{2N}{\sqrt{N^2-1}}\, \oct{F}\, t^a_{jj'} t^a_{kk'}
   \Bigr] \,,
\end{align}
where $j,j',k,k'$ are color indices and $N$ is the number of colors.  For
brevity we omit the labels $a_1, a_2$ on $F$ in this section.  $\sing{F}$
describes the case where the quark lines with equal momentum fractions
$x_1$ or $x_2$ are coupled to a color singlet, whereas in $\oct{F}$ they
form color triplets.  Obviously, a probability interpretation is only
possible for $\sing{F}$, whereas $\oct{F}$ may be regarded as an
interference term in color space.

The color structure of $F_{jj', kk'}$ can alternatively be parameterized
by $\sing{F}$ and the matrix element
\begin{align}
  \label{skewed-color}
\delta_{j'k}\, \delta_{k'j}\, F_{jj', kk'}
&= \mat{(\bar{q}_{3,j}\ms \Gamma_{a_2}\ms q_{2,k})\,
        (\bar{q}_{4,k}\ms \Gamma_{a_1}\ms q_{1,j})}
\nonumber \\
&=
 \frac{\sqrt{N^2-1}}{N}\, \oct{F} + \frac{1}{N}\, \sing{F} \,,
\end{align}
where quark lines with different longitudinal momenta are coupled to color
singlets.  We note that \eqref{skewed-color} becomes equal to $\oct{F}$ in
the large-$N$ limit, except if one has $|\oct{F}| \ll |\sing{F}|$.  By a
Fierz transform one can express \eqref{skewed-color} in terms of matrix
elements
\begin{align}
  \label{skewed-mom}
\sing{\tilde{F}} &=
\mat{(\bar{q}_{4,k}\ms \Gamma_{a_2}\ms q_{2,k})\,
     (\bar{q}_{3,j}\ms \Gamma_{a_1}\ms q_{1,j})} \,,
\end{align}
where the bilinear quark operators have no uncontracted color or spinor
indices.

To illustrate that the color octet combination $\oct{F}$ need not be small
let us consider a three-quark system, as it is done in constituent quark
models.  Since the color part of the three-quark wave function is
$\epsilon_{jkl}$, the color structure of a two-quark distribution then
reads
\begin{align}
F_{jj',kk'} &\;\propto\; 
   \epsilon_{jkl}\, \epsilon_{j'k'l}
 = \delta_{jj'}\, \delta_{kk'} - \delta_{jk'}\, \delta_{kj'} \,,
\end{align}
where $l$ is the color index of the spectator quark and therefore summed
over.  From this one readily obtains $\oct{F} = - \sqrt{2}\, (\sing{F})$,
whereas the combination in \eqref{skewed-color} is found to be equal to $-
\, (\sing{F})$.

With a suitable assignment of color indices, decompositions analogous to
\eqref{quark-color-decomp} can be written down for distributions $F$
involving one or two antiquarks instead of quarks, and for the
interference distributions $I$ in \eqref{interf-dist}.  Given the
normalization of $\oct{F}$ chosen in \eqref{quark-color-decomp}, color
singlet and color octet distributions enter with equal weight
\begin{align}
 \bigl(\ms \sing{F}\ms \sing{F}
         + \oct{F}\ms  \oct{F} \,\bigr) \big/ N^2
\end{align}
in the cross section if each hard scatter produces an electroweak gauge
boson or any other color-singlet system.

Two-gluon distributions have a more involved color
structure.  We first couple each of the gluon pairs $\{ 1,4 \}$ and $\{
2,3 \}$ to an irreducible representation and then couple the two pairs to
an overall color singlet.  This gives
\begin{align}
  \label{gluon-color-decomp}
& F^{aa',bb'} = 
\frac{1}{x_1 p^+\ms x_2\ms p^+}\,
  \mat{(G^{b'}_3\ms \Pi_{a_2}\ms G^{b}_2)\,
       (G^{a'}_4\ms \Pi_{a_1}\ms G^{a}_1)}
\nonumber \\
&= \frac{1}{(N^2-1)^2}\, \Bigl[ \sing{F}\, \delta^{aa'}\ms \delta^{bb'}
   - \frac{\sqrt{N^2-1}}{N}\, \octA{F}\, f^{caa'}\, f^{cbb'}
\nonumber \\
&\quad + \frac{N \sqrt{N^2-1}}{N^2-4}\, \octS{F}\, d^{caa'}\, d^{cbb'}
   + \,\cdots\, \Bigr] \,,
\end{align}
where we use a shorthand notation $G^{b'} \Pi_{a} G^{b} = \Pi_{a}^{jj'}
G^{+j'\!\!,\ms b'}\ms G^{+j,\ms b}$ for the contraction of Lorentz
indices.  Each of the pairs $\{ 1,4 \}$ and $\{ 2,3 \}$ is coupled to a
singlet, an antisymmetric and a symmetric octet in $\sing{F}$, $\octA{F}$
and $\octS{F}$, respectively.  The ellipsis in \eqref{gluon-color-decomp}
stands for terms where the pairs are in higher representations, which are
$10$, $\overline{10}$ and $27$ for $SU(3)$.  The corresponding tensors in
color space can be found in \cite{Mekhfi:1985dv}, cf.\ also App.~A of
\cite{Bartels:1993ih}.  In hard-scattering processes that produce color
singlet states, the distributions enter as
\begin{align}
\bigl(\ms \sing{F}\, \sing{F} + \octA{F}\,  \octA{F}
  + \octS{F}\,  \octS{F} + \cdots \,\bigr) \big/ (N^2-1)^2 \,.
\end{align}
Of course there are also mixed quark-gluon distributions.  The quark lines
can only couple to a color singlet or octet, which has to be matched by
the gluon lines in order to obtain an overall singlet.  A complete
decomposition is thus given by
\begin{align}
  \label{qg-color-decomp}
F_{jj'}^{aa'} &= \frac{1}{x_1 p^+}\,
  \mat{(\bar{q}_{3,j'}\ms \Gamma_{a_2}\ms q_{2,j})\,
       (G^{a'}_4\ms \Pi_{a_1}\ms G^{a}_1)}
\nonumber \\
&= \frac{1}{N (N^2-1)}\, \Bigl[ \sing{F}\, \delta^{aa'}\ms \delta_{jj'}
   - \octA{F}\, \sqrt{2}\, i f^{caa'}\, t^{c}_{jj'}
\nonumber \\
&\quad + \sqrt{\frac{2N^2}{N^2-4}}\, \octS{F}\, d^{caa'}\, t^{c}_{jj'}
  \Bigr] \,.
\end{align}
We note that the color structure of two-parton distributions has already
been discussed in \cite{Mekhfi:1985dv}, using a basis where the parton
pairs $\{ 1,2 \}$ and $\{ 3,4 \}$ are coupled to states of definite color.


\section{Wilson lines}
\label{sec:wilson}

Our discussion so far has been concerned with tree graphs as in
Fig.~\ref{fig:scatter}.  At this level, our results can readily be
generalized to other hard-scattering processes, in particular to jet
production with the well-known subprocesses $qq \to qq$, $qg \to qg$, etc.

A proper factorization formula in QCD must of course include corrections
to the tree-level cross section, and in particular take care of additional
gluon exchange.  Detailed studies of factorization in terms of transverse
momentum dependent parton distributions have been performed for single
Drell-Yan production and its crossed counterparts, $e^+e^-$ annihilation
into back-to-back hadrons and semi-inclusive deep inelastic scattering
\cite{Collins:1981uk,Ji:2004wu,Collins:2007ph, Collins:2011}.  In
\cite{Diehl:2011} we argue that the factorization proof for Drell-Yan
production can to a large part be extended to double hard-scattering
processes producing uncolored states such as electroweak gauge bosons.  We
restrict ourselves to such processes from now on.  For hard-scattering
processes like jet production, serious problems for establishing
transverse-momentum dependent factorization have been encountered even for
single hard scattering \cite{Rogers:2010dm}.  A treatment of the
multiple-scattering case will probably have to wait until this situation
is clarified.

\begin{figure}
\begin{center}
\includegraphics[width=0.35\textwidth]{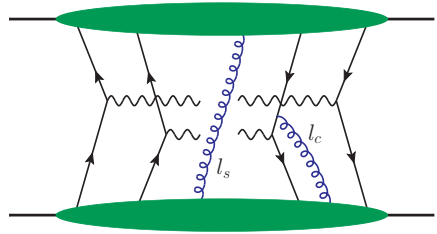}
\end{center}
\caption{\label{fig:extra-gluons} Example graph with a collinear gluon
  $l_c$ and a soft gluon $l_s$.}
\end{figure}

Starting from tree-level diagrams, there are two types of additional gluon
exchange that are not power suppressed in the large scale and hence need
to be taken into account systematically.  The fist type concerns gluons
which emerge from the subgraph representing the partons in the
right-moving proton and which attach to a hard-scattering subgraph, see
Fig.~\ref{fig:extra-gluons}.  These gluons move fast to the right, and to
avoid power suppression their polarization must be in the plus-direction.
To leading-power accuracy, the effect of these gluons can be represented
by Wilson lines that appear in the operators defining parton distributions
and make them gauge invariant.  For gauge boson pair production, each
quark or antiquark field in \eqref{F-position} is to be multiplied by a
Wilson line according to
\begin{align}
q_j(z) &\to [\ms W(z,v) \ms]_{jk}\, q_k(z) \,,
\nonumber \\[0.1em]
\bar{q}_j(z)
  &\to \bar{q}_k(z)\, [\ms W^\dagger(z,v) \ms]_{kj}
\end{align}
with
\begin{align}
  \label{Wilson}
W(z,v) &=  \operatorname{P} \exp\biggl[\; ig \int_0^{\infty} \!
               d\lambda\; v A^a(z - \lambda v)\, t^a \biggr] \,,
\end{align}
where $j$ and $k$ are color indices, $\operatorname{P}$ denotes path
ordering, and the sign convention for the coupling $g$ is such that the
covariant derivative reads $D^\mu = \partial^\mu + i g A^{\mu,a}\ms t^a$.
An analogous discussion holds for left-moving gluons in the left-moving
proton.

One naively expects the vector $v$ to point in the minus direction for
right-moving and in the plus direction for left-moving partons.  This
leads however to rapidity divergences in the parton distributions, due to
contributions from left-moving gluons in a right-moving hadron and vice
versa.  To exclude this unwanted kinematic region and to remove the
divergence, one can take $v$ with nonzero plus- and minus components
\cite{Collins:1981uk,Collins:2007ph}.  This results in an additional
parameter
\begin{align}
  \label{zeta-def}
\zeta^2 &= \frac{(2 p\ms v)^2}{|v^2|}
\end{align}
in the parton distributions for proton $p$.  Their $\zeta$ dependence is
connected with Sudakov logarithms and will be discussed in
Section~\ref{sec:sudakov}.  An adequate choice of $\zeta$ in cross section
formulae is the hard scale $Q$.

The second type of unsuppressed gluon exchange is between the right- and
left-moving partons as shown in Fig.~\ref{fig:extra-gluons}, provided the
gluons are soft and thus do not take partons far off shell.  In processes
with small observed transverse momenta in the final state, the effect of
these gluons does not cancel, but it can be described by a so-called soft
factor, which is defined in terms of vacuum expectation values of Wilson
lines.  Proper care needs to be taken to prevent double counting, because
the Wilson lines $W(z,v)$ in the parton distributions include soft gluon
momenta as well \cite{Ji:2004wu,Collins:2007ph, Collins:2011}.

We now turn to collinear multiparton distributions.  It is well known that
for single-parton distributions the rapidity divergences just mentioned
cancel between real and virtual graphs \cite{Collins:2003fm}, so that in
this case the direction $v$ can be taken exactly lightlike.  The same
argument holds for two-parton distributions in the color singlet channel.
As can be seen from \eqref{F-coll} and \eqref{F-position}, the transverse
separation $\tvec{z}_i$ of quark and antiquark fields is zero in the
collinear distribution $F(x_i, \tvec{y})$.  For color singlet
distributions $\sing{F}$, the Wilson lines whose color indices are
contracted therefore combine to a Wilson line of finite length, going from
$y - \half z_1$ to $y + \half z_1$ or from $-\half z_2$ to $\half z_2$
along the light cone.

The same is, however, not true for color octet distributions $\oct{F}$.
Rapidity divergences from real and virtual graphs do not cancel in that
case, because compared to $\sing{F}$ some graphs change their color
factors whereas others do not.  One must hence keep $v$ away from the
light cone even in collinear octet distributions.  Furthermore, the color
indices of Wilson lines with equal transverse positions do not match, so
that these Wilson lines cannot be combined to a line of finite length.  As
a consequence, terms with color octet distributions in collinear
factorization formulae will have a very different structure from the usual
one.

It should be possible to extend the previous discussion to gluon
distributions.  For the Wilson lines in single-gluon densities a detailed
analysis can be found in \cite{Bomhof:2006dp}, whereas to our knowledge
the soft-gluon sector has not been elaborated.


\section{Parton spin correlations}
\label{sec:spin}

The cross section formulae in Section~\ref{sec:tree} have a nontrivial
dependence on parton spin, even though they are for unpolarized colliding
protons.  $F_{q,q}$ and its counterparts for antiquarks describe
unpolarized partons, whereas all other distributions $F_{\Delta q, \Delta
  q}$, $F_{q,\Delta q}$ etc.\ describe correlations between the
polarization of the two partons, or between their polarization and the
transverse vectors $\tvec{y}$ and $\tvec{k}_i$.  Such spin correlations
have been pointed out in \cite{Mekhfi:1985dv}, but to our knowledge they
have not been implemented in phenomenology.  We find that 16 scalar or
pseudoscalar
functions are required to parameterize the spin structure of $F_{a_1,
  a_2}(x_i, \tvec{k}_i, \tvec{y})$ for a given quark flavor combination.
Eight scalar functions remain when one integrates over $\tvec{k}_1$ and
$\tvec{k}_2$ to obtain $F_{a_1, a_2}(x_i, \tvec{y})$.

To see that parton spin correlations in the proton need not be small, let
us consider a $SU(6)$ symmetric three-quark wave function.  As is well
known, this gives $\Delta u /u = 2/3$ and $\Delta d /d = -1/3$ for the
average longitudinal polarization of $u$ and $d$ quarks, which reproduces
the trend observed in polarized quark densities at $x$ around $0.3$.
Similarly, one obtains $F_{\Delta u, \Delta u} / F_{u, u} = 1/3$ and
$F_{\Delta u, \Delta d} / F_{u, d} = -2/3$ and thus a considerable
correlation between the longitudinal polarization of two quarks.

Parton momentum fractions in processes at LHC are typically well below
$0.3$, say of order $10^{-2}$ or smaller.  In that region polarized
single-parton distributions (to the extent that they are known) are small
compared with their unpolarized counterparts.  This means that there is
little correlation between the respective polarizations of a parton and
the proton when they are far apart in phase space.  From this we should
however not conclude that distributions like $F_{\Delta q, \Delta q}$ are
unimportant at small but comparable $x_1 \sim x_2$, since they describe a
correlation between two partons that are relatively close in phase space.

Parton spin correlations have important consequences in multiple
interaction processes.  Let us again consider gauge boson pair production.
For longitudinal polarization of both the quark and the antiquark that
annihilate into a boson, each spin projector $\Gamma_{\Delta q}$ and
$\Gamma_{\Delta\bar{q}}$ in the hard-scattering cross section
\eqref{quark-X-sect} contains a $\gamma_5$.  After anticommuting one
$\gamma_5$ through the fermion trace, one obtains a factor $\gamma_5^2 =
1$, so that the cross section depends on the combination $\smash{F_{q,q}\,
  F_{\bar{q},\bar{q}} + F_{\Delta q, \Delta q}\, F_{\Delta\bar{q},
    \Delta\bar{q}}}$ of two-parton distributions.  Longitudinal spin
correlations between quarks and antiquarks thus directly affect the rate
of gauge boson pair production by multiple scattering.

The distribution $\smash{F_{\delta q,\delta q}^{jj'}}$ gives rise to even
more striking effects.  Its decomposition into scalar functions contains a
term with $\delta^{jj'}$, which describes a correlation proportional to
the product $\tvec{s}_1 \tvec{s}_2$ of the two transverse quark
polarization vectors.  Recall that the quark field bilinears $\bar{q}
i\sigma^{+j} \gamma_5\ms q$ defining the distribution $\smash{F_{\delta
    q,\delta q}^{jj'}}$ are chiral odd, which corresponds to quarks with
opposite helicities in the amplitude and its conjugate.  As a result,
transverse quark polarization does not contribute to the production of $W$
bosons.  There is however a term with $\smash{F_{\delta q,\delta
    q}^{jj'}\, F_{\delta\bar{q},\delta\bar{q}}^{kk'}}$ in the cross
section for producing two neutral gauge bosons, $\gamma^*\gamma^*$,
$\gamma^* Z$ or $ZZ$.  This term gives in particular rise to an angular
correlation proportional to $\cos 2\varphi$, where $\varphi$ is the
azimuthal angle between the momenta of the negatively charged leptons or
the quarks from the boson decays.  We thus obtain the important result
that correlations between transverse quark and antiquark polarizations
give rise to a correlation between the decay planes of the two produced
bosons.  Moreover, this type of correlation does not arise if the two
bosons are produced by a single hard scattering such as in
Fig.~\ref{fig:scatter}b, where for $q_T \ll Q$ one obtains an angular
modulation with $\cos\varphi$ but none with $\cos 2\varphi$.

It is natural to expect that spin correlations between partons in the
proton also result in correlations between the scattering planes in other
double-scattering processes, such as the production of two dijets.  We
note that in \cite{Berger:2009cm} the absence of such correlations was
explicitly assumed, which in view of our discussion may not be adequate.
It would also be interesting to study whether polarization induced angular
correlations can contribute to the so-called ``ridge effect'' observed in
$pp$ collisions by CMS \cite{Khachatryan:2010gv}.


\section{Mellin moments}
\label{sec:mellin}

If one takes Mellin moments $\int dx_1^{}\, x_1^{n_1} \int dx_2^{}\,
x_2^{n_2}\, (\sing{F})$ of color singlet distributions, the light-cone
operators $\mathcal{O}_{a}(y,z)$ in \eqref{quark-bilinears} turn into
local twist-two operators that are well known from the operator product
expansion.  (As follows from the discussion in Section~\ref{sec:wilson},
the corresponding operators for color octet distributions still contain
Wilson lines and will not be discussed here.)

Let us consider two unpolarized quarks and take the lowest Mellin moment
in both $x_1$ and $x_2$,
\begin{align}
  \label{mellin-moment}
M_{q,q}(\tvec{y}^2) &=
  \int_0^1 dx_1\, \int_0^1 dx_2\,
  \Bigl[ \sing{F}_{q,q}(x_1,x_2, \tvec{y})
\nonumber \\[0.2em]
&\quad
 - \sing{F}_{\bar{q},q}(x_1,x_2, \tvec{y})
 - \sing{F}_{q,\bar{q}}(x_1,x_2, \tvec{y})
\nonumber \\[0.2em]
&\quad
 + \sing{F}_{\bar{q},\bar{q}}(x_1,x_2, \tvec{y}) \Bigr]
\nonumber \\
&= \frac{2}{p^+} \int dy^- 
   \big\langle p \big|\, \mathcal{O}_{q}(0,0)\,
        \mathcal{O}_{q}(y,0) \big| p \big\rangle \,.
\end{align}
We introduce the local operator $\mathcal{O}^\mu(y) = \bar{q}(y)
\gamma^\mu q(y)$ and decompose
\begin{align}
  \label{covar-decomp}
\hspace{-0.5em} \big\langle p \big|\,
   \mathcal{O}^{\nu}\bs (0)\, \mathcal{O}^{\mu}(y) \big| p \big\rangle
&= 2 p^{\mu} p^{\nu}\,
   \langle \mathcal{O}\ms \mathcal{O} \rangle(py, y^2)
 + \ms\cdots \,,
\end{align}
where the ellipsis represents terms with uncontracted vectors $y^{\mu}$,
$y^{\nu}$ or the metric tensor $g^{\mu\nu}$.  The reduced matrix element
$\langle \mathcal{O}\ms \mathcal{O} \rangle$ can only depend on the
invariants $py$ and $y^2$.  We now choose a frame where $\tvec{p} =
\tvec{0}$ and $y^+ = 0$, so that $py = p^+ y^-$ and $y^2 = -\tvec{y}^2$.
Using $\mathcal{O}_q(y,0) = \half \mathcal{O}^+(y) \,|_{y^+ = 0}$ we can
then write \eqref{mellin-moment} in a manifestly covariant form
\begin{align}
  \label{mellin-covar}
M_{q,q}(\tvec{y}^2) &= \int d(py)\;
  \langle \mathcal{O}\ms \mathcal{O} \rangle(py, y^2) 
\,\Big|_{y^2 = -\tvec{y}^2} \,.
\end{align}
It is straightforward to generalize this procedure to higher Mellin
moments and to the other quark polarizations $\Delta q$ and $\delta q$.
In each case one can express the Mellin moment in terms of the matrix
element of a product of two local twist-two operators.

The matrix element in \eqref{mellin-covar} can be evaluated on a lattice
in Euclidean spacetime if one takes $y^0=0$.  This is rather similar to
recent lattice studies of transverse-momentum dependent single-quark
distributions \cite{Hagler:2009mb,Musch:2010ka}.  The restriction to
$y^0=0$ entails
\begin{align}
(py)^2 /(-y^2) &=
   (\mvec{p} \mvec{y})^2 /\mvec{y}^2  \;\le\; \mvec{p}^2 \,,
\end{align}
where $\mvec{p}$ and $\mvec{y}$ denote the spacelike three-vectors.
Results from a discrete Euclidean lattice are hence not sufficient for
evaluating the integral over all $py$ at fixed $y^2$ in
\eqref{mellin-covar}.  This is completely analogous to the situation
discussed in \cite{Musch:2010ka}.  Despite this limitation of principle,
we hope that lattice data in a certain range of $py$ and $y^2$ will one
day provide genuinely nonperturbative information about the behavior of
multiparton distributions.


\section{Connection with generalized parton distributions}
\label{sec:gpds}

To develop a viable phenomenology of multiple interactions, one needs a
simple ansatz for multiparton distributions that can be progressively
refined.  For the distributions which admit a probability interpretation,
a natural starting point is to replace them by the product of
single-parton densities.

To formalize and extend this ansatz, we insert a complete set of
intermediate states $|X\rangle \langle X|$ between the operators
$\mathcal{O}_{a_2}$ and $\mathcal{O}_{a_1}$ in the definition
\eqref{F-impact} of two-parton distributions in impact parameter space.
This gives a product of single-parton operators sandwiched between a
proton state and $X$.  If we \emph{assume} that the ground state dominates
in the sum over all $X$ and take the intermediate proton states in the
impact parameter representation \eqref{impact-states}, then $F(x_i,
\tvec{z}_i, \tvec{y}; \tvec{b})$ involves a product of single-proton
matrix elements
\begin{align}
  \label{product}
& \langle p^+, -\tvec{b} -\half\tvec{d} \,|\ms
  \mathcal{O}_{a_2}(0, z_2) |\, p'^+, \tvec{b}' \rangle
\nonumber \\
& \quad \times \langle p'^+, \tvec{b}' \,|\ms
  \mathcal{O}_{a_1}(y, z_1) |\, p^+, -\tvec{b} +\half\tvec{d} \rangle
\nonumber \\
&= \langle p^+, -\tvec{b} -\half\tvec{d} \,|\ms
  \mathcal{O}_{a_2}(0, z_2) |\, p'^+, \tvec{b}' \rangle\,
  e^{i y^- (p' - p)^+}
\nonumber \\
& \quad \times
  \langle p'^+, \tvec{b}' -\tvec{y} \,|\ms \mathcal{O}_{a_1}(0, z_1)
  |\, p^+, -\tvec{b} -\tvec{y} +\half\tvec{d} \rangle \,.
\end{align}
Integrating the phase factor $e^{i y^- (p' - p)^+}$ over $y^-$ sets $p'^+
= p^+$ in \eqref{F-impact}, and we obtain a representation
\begin{align}
  \label{impact-product}
& \sing{F}_{a_1,a_2}(x_i, \tvec{z}_i, \tvec{y}; \tvec{b})
 \approx f_{a_2}(x_2, \tvec{z}_2; \tvec{b} + \half x_1 \tvec{z}_1)
\nonumber \\[0.2em]
& \quad\qquad\qquad \times
    f_{a_1}(x_1, \tvec{z}_1; \tvec{b} + \tvec{y} - \half x_2 \tvec{z}_2)
\end{align}
in terms of single-quark distributions
\begin{align}
& f_{a}(x, \tvec{z}; \tvec{b}) = \int
  \frac{d^2\tvec{\Delta}}{(2\pi)^2}\; e^{-i \tvec{b} \tvec{\Delta}}
  \int \frac{dz^-}{2\pi}\, e^{i x z^- p^+}
\nonumber \\
&\quad \times 
  \big\langle p^+, \half\tvec{\Delta} \big|
  \mathcal{O}_a(0, z) \big| p^+, - \half\tvec{\Delta} \big\rangle
\end{align}
in impact parameter space.  These distributions satisfy
\begin{align}
& \int \frac{dz^-}{2\pi}\, e^{i x z^- p^+}
  \big\langle p^+, - \tvec{b} - \half\tvec{d} \,\big|
    \mathcal{O}_{a}(0, z) \big| p^+, - \tvec{b} + \half\tvec{d}
  \,\big\rangle
\nonumber \\
 &\quad = \delta^{(2)}\bigl( \tvec{d} - x \tvec{z} \bigr)\,
  f_{a}(x, \tvec{z}; \tvec{b})
\end{align}
in analogy to \eqref{F-impact}.  For $\tvec{z} = \tvec{0}$ they are the
impact parameter dependent parton densities $f_a(x; \tvec{b})$ discussed
in \cite{Soper:1976jc,Burkardt:2002hr} and give the probability to find a
parton with momentum fraction $x$ and distance $\tvec{b}$ from the proton
center.

\begin{figure*}
\begin{center}
\includegraphics[width=0.99\textwidth]{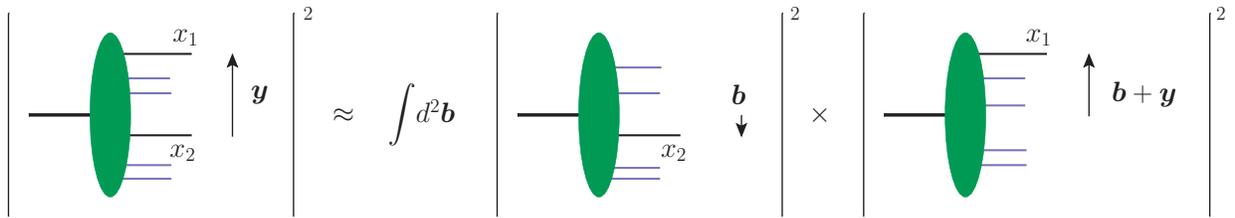}
\end{center}
\caption{\label{fig:impact-reduct} Illustration of the approximation
  \protect\eqref{impact-prod-coll} of a two-parton distribution in terms
  of single-parton distributions.  Implicit in the figure is the
  representation of these distributions as squares of light-cone wave
  functions in the proton.}
\end{figure*}

At $\tvec{z}_i = \tvec{0}$ the relation \eqref{impact-product} involves
only collinear distributions.  Integrating over $\tvec{b}$ we get
\begin{align}
  \label{impact-prod-coll}
\sing{F}_{a_1,a_2}(x_i, \tvec{y}) \approx \int d^2\tvec{b}\;
  f_{a_2}(x_2; \tvec{b}) \, f_{a_1}(x_1; \tvec{b} + \tvec{y}) \,,
\end{align}
which has a straightforward physical interpretation as sketched in
Fig.~\ref{fig:impact-reduct}.  In different guises, this relation is at
the basis of most phenomenological studies and has long been used in the
literature, see e.g.\
\cite{Sjostrand:1986ep,Durand:1987yv,Ametller:1987ru} and
\cite{Frankfurt:2003td,Domdey:2009bg}.  As was noticed in
\cite{Blok:2010ge}, the convolution in \eqref{impact-prod-coll} simplifies
to a product if one Fourier transforms to transverse momentum space, where
one has
\begin{align}
  \label{impact-prod-mom}
\sing{F}_{a_1,a_2}(x_i, \tvec{r}) 
\,\approx\,
  f_{a_2}(x_2; -\tvec{r}) \, f_{a_1}(x_1; \tvec{r})
\end{align}
with $f_{a}(x; \tvec{\Delta}) = \int d^2\tvec{b}\; e^{i \tvec{b}
  \tvec{\Delta}} f_{a}(x; \tvec{b})$.

In the previous argument we have neglected the spin of the proton.  When
inserting intermediate proton states between $\mathcal{O}_{a_2}$ and
$\mathcal{O}_{a_1} $ in a two-parton distribution for an unpolarized
proton, we schematically have
\begin{align}
  \label{proton-helicity-sum}
& \frac{1}{2}\ms \sum_{\lambda}
  \big\langle p, \lambda \,\big|\,
    \mathcal{O}_{a_2}\ms \mathcal{O}_{a_1}
    \,\big|\, p, \lambda \big\rangle
  \,\approx\, \int \frac{dp'^+\, d^2\tvec{p}'}{(2\pi)^3\, 2 p'^+}\,
\nonumber \\
&\quad \times
\frac{1}{2}\ms \sum_{\lambda,\lambda'}
  \big\langle p, \lambda \,\big|\, \mathcal{O}_{a_2}
    \,\big|\, p', \lambda' \big\rangle\,
  \big\langle p', \lambda' \,\big|\, \mathcal{O}_{a_1}
    \,\big|\, p, \lambda \big\rangle
\end{align}
with proton helicities $\lambda$ and $\lambda'$.  The sum on the r.h.s.\
includes matrix elements where the proton helicity differs in the bra and
the ket state.  As an example let us consider the collinear distribution
for two unpolarized quarks.  We then find
\begin{align}
  \label{H-E-result}
& \sing{F}_{q,q}(x_i, \tvec{r}) \,\approx\,
   H^q(x_2, 0, -\tvec{r}^2)\, H^q(x_1, 0, -\tvec{r}^2)
\nonumber \\
&\qquad\quad
 + \frac{\tvec{r}^2}{4 m_p^2}\,
   E^q(x_2, 0, -\tvec{r}^2)\, E^q(x_1, 0, -\tvec{r}^2) \,,
\end{align}
where the first term corresponds to $\lambda=\lambda'$ and the second one
to $\lambda= -\lambda'$ in \eqref{proton-helicity-sum}.  The generalized
parton distributions (GPDs) $H^q(x,\xi,t)$ and $E^q(x,\xi,t)$ can be
studied in hard exclusive processes such as deeply virtual Compton
scattering or vector meson production.  For their definitions and further
information we refer to \cite{Diehl:2003ny,Belitsky:2005qn}.

We emphasize that the relations \eqref{impact-product},
\eqref{impact-prod-coll} and \eqref{impact-prod-mom} are obtained by
restricting a sum over all intermediate states to a single proton in a
selected helicity state.  We do not have a justification or a strong
physical motivation for this restriction, other than stating a posteriori
that it is tantamount to neglecting any correlation between two partons in
the proton.  It seems plausible to assume that this is a reasonable first
approximation, at least in a certain region of variables, but one should
not expect such an approximation to be very precise.  The result
\eqref{H-E-result} already goes beyond the complete neglect of
correlations, and the relative size of the term depending on $e^q$ may be
taken as an indicator for the level of precision of the simplest
factorized ansatz.  Possible deviations from \eqref{impact-prod-coll} and
their consequences for multiple scattering cross sections have e.g.\ been
discussed in \cite{Domdey:2009bg,Calucci:1997ii,Rogers:2009ke}.

So far we have discussed the distributions $\sing{F}_{a_1,a_2}$, which can
be interpreted as probabilities or pseudo-probabilities.  The formal
derivation we have sketched can however be extended to distributions that
have interference character.  Repeating the above steps for the
distributions appearing in Fig.~\ref{fig:interference}b, one obtains
matrix elements of operators $\bar{d} \ms\Gamma\ms u$ or $\bar{u}
\ms\Gamma\ms d$ that transfer isospin.  For a proton target, the ground
state among the intermediate states inserted between the two operators is
then a neutron.  Isospin symmetry relates the resulting matrix elements as
$\langle n | \bar{d}\, \Gamma\ms u | p \rangle = \langle p | \bar{u}\ms
\Gamma\ms d | n \rangle = \langle p | \bar{u}\ms \Gamma\ms u | p \rangle -
\langle p | \bar{d}\, \Gamma\ms d | p \rangle$.  Similar relations for
strange quarks assume flavor $SU(3)$ symmetry and can be found in
\cite{Diehl:2003ny,Belitsky:2005qn}.

Inserting physical intermediate states is useful between the color singlet
operators appearing in $\sing{F}$ but not between the color octet
operators $\bar{q}\ms \Gamma\ms t^a q$ in $\oct{F}$.  One can however
repeat the preceding construction for the distributions $\sing{\tilde{F}}$
defined in \eqref{skewed-mom}.  In that case the two partons represented
by a color singlet operator carry different plus-momentum fractions $x_1$
and $x_2$.  This implies $p'^+ \neq p^+$ in the resulting product of
proton matrix elements, which are therefore associated with GPDs at
nonzero skewness $\xi = {}\pm (p-p')^+ /(p+p')^+$.  Likewise, the
rearrangement of fields to color singlet operators in the interference
distribution \eqref{interf-dist} leads to GPDs with nonzero $\xi$.  We
note that in the exclusive processes where GPDs can be measured, one
always has $\xi\neq 0$ because the scattered proton must have a smaller
momentum than the proton target.

GPDs give rather direct information about the distribution of partons in
impact parameter $\tvec{b}$, which is Fourier conjugate to a measurable
transverse momentum $\tvec{\Delta}$ in physical processes.  By contrast,
the interparton distance $\tvec{y}$ in multiple interactions is integrated
over in the cross section formula \eqref{X-sect-mixed} and not directly
related to any measurable kinematic distribution.  Although the connection
between GPDs and multiparton distributions is not exact, it provides an
opportunity to obtain some quantitative information about impact parameter
dependence from an independent source.  An important example is the
correlation between impact parameter and longitudinal momentum of partons,
for which there are clear indications in GPDs (see Section~4.4.5 of
\cite{Hagler:2009ni} and references therein) and which has recently been
studied for multiparton interactions in \cite{Corke:2011yy}.


\section{Ladder graphs}
\label{sec:ladders}

The factorization formulae in Section~\ref{sec:tree} are for kinematics
where the transverse boson momenta are much smaller than the large scale
$Q$.  This includes but is not limited to the region where $q_T$ is
comparable to a hadronic scale $\Lambda$.  In this and the following
section we consider the region of intermediate transverse momenta $\Lambda
\ll q_T \ll Q$.  If $\tvec{q}_1$ and $\tvec{q}_2$ are large compared with
$\Lambda$ then at least some of the transverse momenta in the two-parton
distributions must be large as well.  This opens the way for a
perturbative description where the transverse-momentum dependent
distributions factorize into collinear distributions and a hard
parton-level scattering.  This increases the predictive power of the
theory, given that collinear two-parton distributions depend on fewer
variables and are less numerous.  Note that the scale $q_T$ of the hard
scattering in the two-parton distributions is still much smaller than the
scale $Q$ of the multiple hard scattering in the proton-proton collision.

\begin{figure}
\begin{center}
\includegraphics[width=0.44\textwidth]{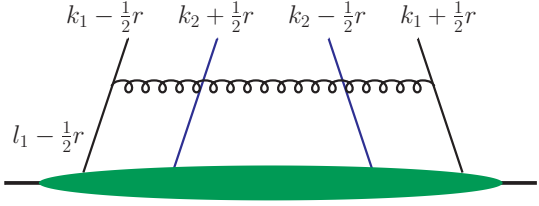}
\end{center}
\caption{\label{fig:ladders} Ladder graph for a two-parton distribution at
  large $\tvec{k}_1$ and small $\tvec{k}_2$ and $\tvec{r}$.}
\end{figure}

The type of graphs that describe the hard scattering in $F(x_i,
\tvec{k}_i, \tvec{r})$ depends on the relative size of the
transverse-momentum arguments.  For instance, the single-ladder graph in
Fig.~\ref{fig:ladders} gives large $\tvec{k}_1$, whereas a double-ladder
graph with an additional gluon exchanged between the quarks with momentum
fraction $x_2$ leads to large $\tvec{k}_1$ and $\tvec{k}_2$.  In both
cases no hard parton is exchanged between the lines with momentum fraction
$x_1$ and those with momentum fraction $x_2$, so that $\tvec{r}$ is of
order $\Lambda$.  This corresponds to an interparton distance $\tvec{y}$
of hadronic size.

The computation of the graph in Fig.~\ref{fig:ladders} proceeds exactly as
for single-quark distributions at large transverse momentum and can e.g.\
be found in \cite{Collins:2011,Ji:2006vf,Bacchetta:2008xw}.  One obtains
\begin{align}
  \label{single-ladder-y}
& \int d^2\tvec{k}_2\; F^J(x_1,x_2, \tvec{k}_1, \tvec{k}_2, \tvec{y})
  \bigg|_{\text{Fig.~\protect\ref{fig:ladders}}}
 = \frac{1}{\pi\, \tvec{k}_1^2}
\nonumber \\
&\quad \times \sum_{J'}
    \int_{x_1}^{1-x_2} \frac{du_1}{u_1}\,
    P^{JJ'}\Bigl( \frac{x_1}{u_1} \Bigr)\,
    F^{J'}(u_1,x_2, \tvec{y}) \,,
\end{align}
where $P$ has a logarithmic dependence on $\tvec{k}_1^2$ which we have not
displayed for brevity.  We have integrated the distribution over
$\tvec{k}_2$ since this turns out to be the quantity that appears in the
cross section at large $\tvec{q}_1$ and $\tvec{q}_2$.  The indices $J$ and
$J'$ indicate that one has a matrix structure if transitions between gluon
and quark distributions are permitted by the quantum numbers.  Quark-gluon
transitions occur for the combinations
\begin{align}
{}^{1\!}F^J &=
\begin{pmatrix}
  \sum_{q} \, [ \sing{F}_{q,a} + \sing{F}_{\bar{q},a} ] \\[0.3em]
  \sing{F}_{g,a}
\end{pmatrix} \,,
\nonumber \\
{}^{S\!}F^{J} &= 
\begin{pmatrix}
  \sum_{q} \, [ \oct{F}_{q,a} + \oct{F}_{\bar{q},a} ] \\[0.3em]
  \octS{F}_{g,a}
\end{pmatrix} \,,
\nonumber \\
{}^{A\!}F^{J} &= 
\begin{pmatrix}
  \sum_{q} \, [ \oct{F}_{q,a} - \oct{F}_{\bar{q},a} ] \\[0.3em]
  \octA{F}_{g,a}
\end{pmatrix} \,,
\end{align}
where the second parton index $a$ can indicate an unpolarized or polarized
quark or antiquark.  If $a$ indicates a gluon, one should replace
$\oct{F}_{q,a} + \oct{F}_{\bar{q},a}$ by $\octS{F}_{q,a} +
\octS{F}_{\bar{q},a}$ and $\oct{F}_{q,a} - \oct{F}_{\bar{q},a}$ by
$\octA{F}_{q,a} - \octA{F}_{\bar{q},a}$.
Taking into account the color factors in the definitions
\eqref{quark-color-decomp}, \eqref{gluon-color-decomp} and
\eqref{qg-color-decomp}, we find splitting matrices
\begin{align}
  \label{splitting-matrices}
{}^{1\!}P^{JJ'} &=
\begin{pmatrix}
C_F P_{qq} & n_F P_{qg} \\[0.3em]
C_F P_{gq} & N P_{gg}
\end{pmatrix} ,
\nonumber \\[0.4em]
{}^{S\!}P^{JJ'} &=
\begin{pmatrix}
-\frac{1}{2N}\ms P_{qq} &
 \sqrt{\frac{N^2-4}{2 (N^2-1)}}\; n_F P_{qg} \\[0.5em]
 \sqrt{\frac{N^2-1}{8}}\ms \sqrt{\frac{N^2-4}{N^2}}\, P_{gq} &
 \frac{N}{2}\ms P_{gg}
\end{pmatrix} ,
\nonumber \\[0.4em]
{}^{A\!}{P}^{JJ'} &=
\begin{pmatrix}
-\frac{1}{2N}\ms P_{qq} &
 \sqrt{\frac{N^2}{2 (N^2-1)}}\; n_F P_{qg} \\[0.5em]
 \sqrt{\frac{N^2-1}{8}}\, P_{gq} &
 \frac{N}{2}\ms P_{gg}
\end{pmatrix}
\end{align}
for $N$ colors and $n_F$ quark flavors, where the functions $P_{qq}(z)$,
$P_{qg}(z)$, etc.\ are defined without color factors and differ from the
usual DGLAP splitting functions at most by terms proportional to
$\delta(1-z)$.  The color factors in \eqref{splitting-matrices} agree with
those in \cite{Bukhvostov:1985rn} if one restores a missing factor
$\sqrt{2/N}$ in the expression of $P_{8f}$, eq.~(54b) of that paper.  This
factor is also required for the normalization of $P_{8f}$ as a projector
in eq.~(53) of~\cite{Bukhvostov:1985rn}.

We see that color factors are always smaller in the octet channels than in
the singlet channel, with the biggest suppression occurring for $P_{qq}$.
In the large-$N$ limit the singlet matrix ${}^{1\!}P$ has eigenvalues $N
P_{gg}$ and $\smash{\frac{N}{2}} P_{qq}$, whereas for both ${}^{S\!}P$ and
${}^{A\!}P$ one of the eigenvalues is $\smash{\frac{N}{2}} P_{gg}$ and the
other is of order $1$.  This suggests a relative suppression of color
octet distributions at large transverse momentum, but quantitative
estimates are needed to assess how important this suppression is.

Quark-gluon transitions do not occur for the combinations $\sum_q \,[
\sing{F}_{q,a} - \sing{F}_{\bar{q},q} ]$, for the difference of
distributions for two quark flavors, and for matrix elements where the
quark flavors differ on the two sides of the final-state cut as in
Fig.~\ref{fig:interference}b.  In these cases the indices $J$ and $J'$ in
\eqref{single-ladder-y} take only one value, with the splitting function
$P$ being $C_F P_{qq}$ for color singlet and $- \frac{1}{2N}\ms P_{qq}$
for color octet combinations.  Furthermore, there is no transition between
gluons and the interference distributions in Fig.~\ref{fig:interference}a.
From the experience with single-parton distributions one can expect that
at low $x_1$ and high $\tvec{k}_1$ those combinations of quark and
antiquark distributions will dominate that receive a contribution from
gluons on the r.h.s.\ of \eqref{single-ladder-y}.

The splitting matrices for longitudinally polarized quarks and gluons
contain different splitting functions $\Delta P_{qq}$, $\Delta P_{qg}$,
etc.\ but have the same color structure as \eqref{splitting-matrices}.  No
transitions occur between $\delta q$, $\delta\bar{q}$ and $\delta g$,
since the former describe parton helicity flip by one unit and latter by
two units.  A relation similar to \eqref{single-ladder-y} can be derived
for double-ladder graphs and involves the same splitting matrices as
above.

We finally note that for $q_T \gg \Lambda$ the contributions from ladder
graphs lead to a power behavior
\begin{align}
  \label{ladders-X-sect}
\frac{s^2\ms d\sigma}{\prod_{i=1}^2 dx_i\, d\bar{x}_i\, d^2\tvec{q}{}_i}
  \,\bigg|_{\text{ladders}}
 &\sim \frac{\Lambda^2}{q_T^4}
\end{align}
of the double-scattering cross section scaled with $s^2$.


\section{Parton splitting}
\label{sec:splitting}

\begin{figure}
\begin{center}
\includegraphics[width=0.40\textwidth]{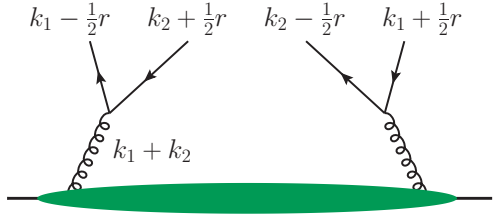}
\end{center}
\caption{\label{fig:high-2-4} Lowest-order graph for a quark-antiquark
  distribution at large $|\tvec{r}| \sim |\tvec{k}_i|$.  The two partons
  with momentum fractions $x_1$ and $x_2$ originate from the splitting of
  a single gluon.}
\end{figure}

\begin{figure}
\begin{center}
\includegraphics[width=0.43\textwidth]{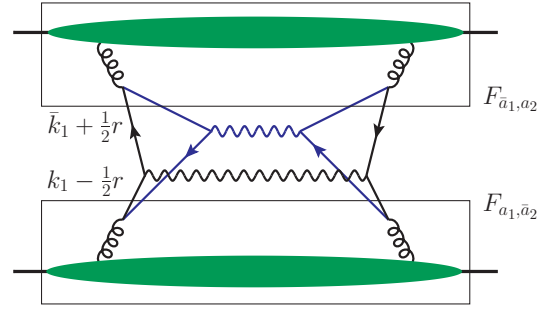}
\end{center}
\caption{\label{fig:high-2-4-X} Graph for the cross section where both
  two-parton distributions $F_{a_1, \bar{a}_2}$ and $F_{\bar{a}_1, a_2}$
  (indicated by boxes) involve the splitting of one into two partons.}
\end{figure}

We now consider the case where $|\tvec{r}| \sim |\tvec{k}_i|$ is large,
which corresponds to perturbatively small distances $|\tvec{y}|$.  Power
counting analysis shows that the dominant contributions to the
double-scattering cross section are from graphs where two partons in the
$t$ channel split into four partons.  A lowest-order example for $F_{a_1,
  \bar{a}_2}$ is shown in Fig.~\ref{fig:high-2-4}.  At next order in
$\alpha_s$ graphs appear where the hard scattering if fully connected,
e.g.\ with an additional gluon exchanged between the left- and right-hand
sides of Fig.~\ref{fig:high-2-4}.  The power behavior of the scaled cross
section is found to be
\begin{align}
  \label{splitting-X-sect}
\frac{s^2\ms d\sigma}{\prod_{i=1}^2 dx_i\, d\bar{x}_i\, d^2\tvec{q}{}_i}
  \,\bigg|_{\text{splitting}}
 &\sim \frac{1}{q_T^2}
\end{align}
if in at least one of the colliding protons the two-parton distribution
has a fully connected hard scattering.  If in both distributions the
hard-scattering subprocess is disconnected as in
Fig.~\ref{fig:high-2-4-X}, the scaled cross section behaves like
$1/\Lambda^2$ instead, but the final-state phase space is then restricted
to $|\tvec{q}_1 + \tvec{q}_2| \sim \Lambda$.

We see that the contribution \eqref{ladders-X-sect} from ladder graphs at
large $\tvec{y}$ is suppressed by $\Lambda^2 /q_T^2$ compared with the
splitting contribution \eqref{splitting-X-sect} at small $\tvec{y}$.  This
suppression may however be compensated by a stronger increase of
\eqref{ladders-X-sect} when $x_1$ and $x_2$ become small.  It is known
that in appropriate quantum number channels the iteration of ladder graphs
leads to a growth of parton densities at small momentum fractions.  The
two-parton distribution at the bottom of Fig.~\ref{fig:ladders} contains
one ladder between the lines with momentum fraction $x_1$ and another
between the lines with momentum fraction $x_2$, whereas at the bottom of
Fig.~\ref{fig:high-2-4} there is a single-parton density containing only
one ladder.  To establish whether the $\Lambda^2 /q_T^2$ suppression or
the small-$x$ enhancement of the ladder-graph contributions is more
important in given kinematics requires a detailed study.

The calculation of the graph in Fig.~\ref{fig:high-2-4} reveals a number
of important issues.  We give here only the contribution from the
transverse-momentum dependent unpolarized gluon distribution
$f_g(x,\tvec{k})$ and omit terms depending on the gluon Boer-Mulders
function \cite{Boer:2010zf}, which describes the correlation between the
transverse momentum and the linear polarization of a gluon.  In the color
singlet channel we find
\begin{align}
  \label{splitting-result}
& \sing{F}_{a_1,\bar{a}_2}(x_i, \tvec{k}_i, \tvec{r})
   \,\Big|_{\text{Fig.~\protect\ref{fig:high-2-4}}}
= \frac{\alpha_s}{4\pi^2}\,
  \frac{f_g(x_1+x_2, \tvec{k}_1 + \tvec{k}_2)}{x_1+x_2}
\nonumber \\
&\quad \times
  T_{a_1,\bar{a}_2}^{l\ms l'}\biggl( \frac{x_1}{x_1+x_2} \biggr)\;
  \frac{\bigl( \tvec{k} + \half \tvec{r} \bigr)^{l}
        \bigl( \tvec{k} - \half \tvec{r} \bigr)^{l'}}{%
           \bigl( \tvec{k} + \half \tvec{r} \bigr)^2
           \bigl( \tvec{k} - \half \tvec{r} \bigr)^2}
\end{align}
with $\tvec{k} = \smash{\half} (\tvec{k}_1 - \tvec{k}_2)$.  For the color
octet distributions $\oct{F}_{a_1,\bar{a}_2}$ we obtain the same
expression times a suppression factor $- 1/\sqrt{N^2-1}$.  We have
\begin{align}
T_{q,\bar{q}}^{l\ms l'}(u)
  = - T_{\Delta q, \Delta\bar{q}}^{l\ms l'}(u)
 &= \delta^{l\ms l'}\, \bigl[\ms u^2 + (1-u)^2 \ms\bigr] \,,
\nonumber \\[0.3em]
T_{\Delta q, \bar{q}}^{l\ms l'}(u)
  = - T_{q, \Delta\bar{q}}^{l\ms l'}(u) \hspace{1.2ex}
 &= i\epsilon^{l\ms l'}\ms \bigl[\ms u^2 - (1-u)^2 \ms\bigr] \,,
\nonumber \\[0.3em]
T_{\delta q, \delta\bar{q}}^{jj',\, l\ms l'}(u)
 &= {}- \delta^{jj'}_{} \delta^{l\ms l'}\, 2 u (1-u) \,,
\end{align}
whereas $T_{a_1, \bar{a}_2} = 0$ for the other polarization combinations.
We see that several nontrivial and large spin correlations are generated
by perturbative splitting.  To which extent these correlations persist at
nonperturbative values of $\tvec{y}$ is an interesting question from the
point of view of hadron structure.

With the analog of \eqref{X-sect-mom} for the product $F_{a_1, \bar{a}_2}
F_{\bar{a}_1, a_2}$ one finds that the contribution of
Fig.~\ref{fig:high-2-4-X} to the cross section for two-boson production is
proportional to
\begin{align}
  \label{splitting-X-mom}
\int d^2\tvec{k}_{+}\,
\frac{\tvec{k}{}_{+}^{\ms l} (\tvec{k}_{+}^{} - \tvec{q})^{m}}{%
      \tvec{k}{}_{+}^2 (\tvec{k}_{+}^{}  - \tvec{q})^2} \;
\int d^2\tvec{k}_{-}\, 
\frac{\tvec{k}{}_{-}^{\ms l'} (\tvec{k}_{-}^{} - \tvec{q})^{m'}}{%
      \tvec{k}{}_{-}^2 (\tvec{k}_{-}^{}  - \tvec{q})^2}
\end{align}
with $\tvec{q} = \smash{\half} (\tvec{q}_1 - \tvec{q}_2)$ and
$\tvec{k}_{\pm} = \tvec{k} \pm \half\tvec{r}$.  Each inte\-gral is
infrared finite but has a logarithmic divergence at large $\tvec{k}_\pm$.
To assess the meaning of these divergences, we first recall that the
derivation of the cross section formula \eqref{X-sect-mom} and its analogs
for other channels assumes transverse parton momenta $|\tvec{k}_i \pm
\half\tvec{r}| \ll Q$.  As it turns out, the integrand in
\eqref{splitting-X-mom} does not decrease fast enough to suppress the
region of large $\tvec{k}_{\pm}$, where the approximations leading to
\eqref{X-sect-mom} are invalid.  On the other hand, we note that
Fig.~\ref{fig:high-2-4-X} can also be read as a graph for producing two
gauge bosons $V_1$ and $V_2$ in a single hard-scattering process.  For the
parton level amplitude it represents a box graph, which indeed diverges
logarithmically in the ultraviolet.  The full amplitude for $g g\to V_1\ms
V_2$ is however finite, because the divergences of all contributing box
graphs cancel each other.  This was noted in \cite{Gaunt:2011xd} and has
long been known for the related process $\gamma\gamma \to \gamma\gamma$
with on-shell photons \cite{Karplus:1950zza}.  A detailed analysis of the
infrared behavior of $g g\to V_1\ms V_2$ is given in \cite{Gaunt:2011xd},
and explicit expressions for the $g g\to Z\ms Z$ amplitude can be found in
\cite{Glover:1988rg}.

Since the definition of two-parton distributions we have used so far
includes the splitting contribution \eqref{splitting-result}, this
definition is not appropriate for the cross section formula
\eqref{X-sect-mom}, and one or both of them needs to be modified.  It
remains a task for future work to devise a consistent formulation that is
also suitable for practical computations.  Such a formulation must
obviously avoid the divergent integrals in \eqref{splitting-X-mom}.  In
addition it must address the problem of double counting in
Fig.~\ref{fig:high-2-4-X} when the contributions from single and double
hard scattering are added in the cross section.  The analogous double
counting problem for multijet production was pointed out in
\cite{Cacciari:2009dp}.


\section{Collinear distributions}
\label{sec:evolution}

We have encountered collinear two-parton distributions in two different
contexts, firstly in the cross section for multiple scattering when
final-state momenta are integrated over, and secondly in the calculation
of two-parton distributions at high transverse momentum.  In this section
we discuss the scale evolution of collinear distributions at leading order
in $\alpha_s$.  We focus on the color singlet sector, which is most
similar to the case of single-parton densities.  (As noted at the end of
Section~\ref{sec:wilson}, collinear color octet distributions involve
further complications related with rapidity divergences.)

Integrating $F(x_i, \tvec{k}_i, \tvec{r})$ over the transverse momenta
$\tvec{k}_i$ gives logarithmic divergences, which is evident from the
factor $1/\tvec{k}_1^2$ in \eqref{single-ladder-y}.  The definition of
collinear distributions requires subtraction of these divergences, in full
analogy to the case of single-parton distributions.  We now discuss the
contribution of the splitting graph in Fig.~\ref{fig:high-2-4} and
consider unpolarized quarks for definiteness.  The expression
\eqref{splitting-result} behaves like $1/ \tvec{k}^2$ for large $\tvec{k}$
and thus gives a logarithmically divergent integral.  This divergence
needs to be subtracted as well.  Notice that the four quark lines in
Fig.~\ref{fig:high-2-4} are far off-shell if \emph{one} of the transverse
momenta $\tvec{r}$ and $\tvec{k}$ is large.  This implies that
\eqref{splitting-result} describes not only the large $\tvec{r}$ behavior
of $F_{a_1,\bar{a}_2}(x_i, \tvec{k}_i, \tvec{r})$ but also its behavior
for large $\tvec{k}$ at small $\tvec{r}$.

Calculating the graphs in Figs.~\ref{fig:ladders} and \ref{fig:high-2-4}
in $4-2\epsilon$ dimensions, performing $\overline{\text{MS}}$ subtraction
of the ultraviolet divergent terms, and adding the contributions from
self-energy graphs, one obtains the evolution equation
\begin{align}
  \label{inhom-dglap}
& \frac{d}{d\log\mu^2}\, \sing{F}_{q, \bar{q}}(x_1,x_2, \tvec{r})
\nonumber \\
 &\quad = \sum_{b_1 = q,g}\,
    \int_{x_1}^{1-x_2} \frac{du_1}{u_1}\,
       P_{q,b_1}\biggl(\frac{x_1}{u_1}\biggr)\,
         \sing{F}_{b_1, \bar{q}}(u_1,x_2, \tvec{r})
\nonumber \\
 &\quad + \sum_{b_2 = \bar{q},g}\,
    \int_{x_2}^{1-x_1} \frac{du_2}{u_2}\,
       P_{\bar{q},b_2}\biggl(\frac{x_2}{u_2}\biggr)\,
         \sing{F}_{q, b_2}(x_1,u_2, \tvec{r})
\nonumber \\
 &\quad + \frac{1}{x_1+x_2}\,
       P_{q,g}\biggl( \frac{x_1}{x_1+x_2} \biggr)\,
       f_g(x_1+x_2) \,,
\end{align}
which is valid for any value of $\tvec{r}$.  Here $P_{b_1, b_2}$ denotes
the usual DGLAP splitting functions (including color factors).
Remarkably, the $\tvec{y}$ dependent distributions $F(x_i, \tvec{y})$
evolve differently.  Working in $4-2\epsilon$ dimensions, one finds that
for nonzero $\tvec{y}$
\begin{align}
& \int \frac{d^{2-2\epsilon}\tvec{r}\;
             d^{2-2\epsilon}\tvec{k}}{(2\pi)^{2-2\epsilon}}\;
        e^{-i \tvec{r} \tvec{y}}\,
  \frac{\bigl( \tvec{k} + \half \tvec{r} \bigr)
        \bigl( \tvec{k} - \half \tvec{r} \bigr)}{%
           \bigl( \tvec{k} + \half \tvec{r} \bigr)^2
           \bigl( \tvec{k} - \half \tvec{r} \bigr)^2}
\end{align}
has the finite value $1/\tvec{y}^2$ at $\epsilon=0$ and hence requires no
ultraviolet subtraction.  One thus obtains a homogeneous evolution
equation
\begin{align}
  \label{hom-dglap}
& \frac{d}{d\log\mu^2}\, \sing{F}_{q, \bar{q}}(x_1,x_2, \tvec{y})
\nonumber \\
 &\quad = \sum_{b_1 = q,g}\,
    \int_{x_1}^{1-x_2} \frac{du_1}{u_1}\,
       P_{q,b_1}\biggl(\frac{x_1}{u_1}\biggr)\,
         \sing{F}_{b_1, \bar{q}}(u_1,x_2, \tvec{y})
\nonumber \\
 &\quad + \sum_{b_2 = \bar{q},g}\,
    \int_{x_2}^{1-x_1} \frac{du_2}{u_2}\,
       P_{\bar{q},b_2}\biggl(\frac{x_2}{u_2}\biggr)\,
         \sing{F}_{q, b_2}(x_1,u_2, \tvec{y}) \,.
\end{align}
This result is very natural if one considers that $F_{a_1, \bar{a}_2}(x_i,
\tvec{y})$ is defined in terms of the product
\begin{align}
  \label{twice-twist-2}
\bigl[ \bar{q}(\half z_2)\, \gamma^+ \ms q(-\half z_2) \bigr]\,
\bigl[ \bar{q}(y-\half z_1)\, \gamma^+\ms q(y+\half z_1) \bigr]
\end{align}
of two operators with light-like field separations $z_1^2 = z_2^2 = 0$,
where we have omitted the Wilson lines between the fields for brevity.
Each of the two operators needs to be renormalized in the same way as in a
single-parton density, but as long as the operators are taken at a finite
spacelike distance $y^2 = -\tvec{y}^2$ no further short-distance
singularities appear.

It has long been known that
\begin{align}
  \label{all-int}
F_{q,\bar{q}}(x_i) &= F_{q,\bar{q}}(x_i,
\tvec{r}=\tvec{0}) 
= \int_{\text{reg}} \!\!\!\! d^2\tvec{y}\; F_{q,\bar{q}}(x_i, \tvec{y})
\end{align}
evolves as in \eqref{inhom-dglap}, see
\cite{Kirschner:1979im,Shelest:1982dg,Snigirev:2003cq} and the recent
studies \cite{Gaunt:2009re,Ceccopieri:2010kg}.  In position space
representation, the inhomogeneous term now appears because the behavior
\begin{align}
  \label{small-y-behavior}
F_{q,\bar{q}}(x_i, \tvec{y}) \sim 1/ \tvec{y}^2
\end{align}
at small $\tvec{y}$ induces a logarithmic divergence in the integral over
$\tvec{y}$, which needs to be regularized as indicated by the subscript
``reg'' in \eqref{all-int}.  We note that $F(x_i)$ does not directly
appear in double-scattering cross sections but is often used as an
intermediate quantity for modeling the distributions $F(x_i, \tvec{y})$.

We have already seen in \eqref{splitting-X-mom} that the contribution of
splitting graphs to the multiparton distributions leads to logarithmic
divergences in the differential cross section.  An even worse divergence
appears in the cross section integrated over $\tvec{q}_1$ and
$\tvec{q}_2$, which according to \eqref{X-sect-mom} and
\eqref{X-sect-mixed} involves an integral
\begin{align}
  \label{X-sect-int}
\int d^2\tvec{y}\,
    F(x_i, \tvec{y}) F(\bar{x}_i, \tvec{y})
= \int \frac{d^2\tvec{r}}{(2\pi)^2}\,
    F(x_i, \tvec{r}) F(\bar{x}_i, -\tvec{r}) \,.
\end{align}
This is linearly divergent in $\tvec{y}^2$ according to
\eqref{small-y-behavior}.  It also diverges linearly in $\tvec{r}^2$,
because for large $\tvec{r}$ one has $F(x_i, \tvec{r}) \sim
\log(\tvec{r}^2/\mu^2)$.  The extra ultraviolet subtraction included in
the definition of $F(x_i, \tvec{r})$ is insufficient to regulate this
divergence in the cross section.  In the recent paper \cite{Ryskin:2011kk}
a finite result is obtained by imposing a cutoff $\tvec{r}^2 < \min(q_1^2,
q_2^2)$ in \eqref{X-sect-int}.

In this context we wish to comment on an ansatz often made in
phenomenological studies, where the $\tvec{y}$ dependent two-parton
distributions are written as
\begin{align}
  \label{factor-ansatz}
F(x_i, \tvec{y}; \mu) = f(\tvec{y})\, F(x_i; \mu)
\end{align}
with a smooth function $f(\tvec{y})$ satisfying the normalization
condition $\int d^2\tvec{y}\, f(\tvec{y}) = 1$.  A typical choice for
$f(\tvec{y})$ is for instance a Gaussian.  This type of ansatz is
obviously inconsistent if $F(x_i, \tvec{y}; \mu)$ is defined from the
product \eqref{twice-twist-2} of twist-two operators, since the $\mu$
dependence of the l.h.s.\ is then given by the homogeneous evolution
equation \eqref{hom-dglap} whereas the $\mu$ dependence of the r.h.s.\ is
governed by an inhomogeneous evolution equation as in \eqref{inhom-dglap}.
If one instead defines the $\tvec{y}$ dependent distribution as the
Fourier transform of $F(x_i, \tvec{r})$, then the ansatz
\eqref{factor-ansatz} is consistent regarding evolution since $F(x_i,
\tvec{r})$ evolves according to \eqref{inhom-dglap}.  However, we do not
think that is a satisfactory definition, since it does not cure the
divergence of the integral \eqref{X-sect-int} in double-scattering cross
sections.  An ansatz like \eqref{factor-ansatz} with a smooth function
$f(\tvec{y})$ does not have this problem and may be regarded as modeling a
$\tvec{y}$ distribution in which the perturbative splitting contribution
giving rise to the $1/\tvec{y}^2$ singularity has been removed.  Since the
ansatz is ad hoc, one cannot say what the correct evolution equation to be
used on both sides of \eqref{factor-ansatz} actually is.  Our discussion
suggests that the homogeneous form \eqref{hom-dglap} may be more
appropriate, at least for values $\tvec{y}$ of typical hadronic size,
which are of course most important when the ansatz is used in the
factorization formula.  A theoretically sound solution remains a task for
future work.


\section{Sudakov logarithms}
\label{sec:sudakov}

As is well known, transverse momenta $\tvec{q}_i$ which are much smaller
than the hard scale $Q$ of a process give rise to Sudakov logarithms in
the cross section.  These logarithms must be resummed to all orders in
perturbation theory, which for single gauge boson production can be done
using the Collins-Soper-Sterman formalism \cite{Collins:1984kg}.  We
extend this formalism to gauge boson pair production in \cite{Diehl:2011}
and sketch the main results of our analysis in the following.

The dependence of a two-quark distribution on the rapidity parameter
$\zeta$ defined in \eqref{zeta-def} is governed by the differential
equation
\begin{align}
  \label{general-cs-eq}
& \frac{d}{d\log\zeta}\, \begin{pmatrix}
  \sing{F} \\[0.2em]
  \oct{F} \end{pmatrix}
 = \bigl[ G(x_1 \zeta, \mu) + G(x_2 \zeta, \mu) + K(\tvec{z}_1, \mu)
\nonumber \\
 &\quad + K(\tvec{z}_2, \mu) \bigr]
    \begin{pmatrix}
      \sing{F} \\[0.2em]
      \oct{F} \end{pmatrix}
+ \mathbf{M}(\tvec{z}_1, \tvec{z}_2, \tvec{y})
      \begin{pmatrix} \sing{F} \\[0.2em]
      \oct{F} \end{pmatrix} \,,
\end{align}
where $\sing{F}$ and $\oct{F}$ depend on $x_i$, $\tvec{z}_i$, $\tvec{y}$
and $\zeta$.  They also depend on a renormalization scale $\mu$, but we
need not discuss this dependence here.  The kernels $G$ and $K$ in
\eqref{general-cs-eq} already appear in the Collins-Soper equation
\cite{Collins:1981uk} for single-quark distributions,
\begin{align}
  \label{cs-single}
\frac{d f(x,\tvec{z}; \zeta)}{d\log\zeta}
&= \bigl[ G(x \zeta, \mu) + K(\tvec{z}, \mu) \bigr]\,
   f(x,\tvec{z}; \zeta) \,.
\end{align}
The matrix $\mathbf{M}$ mixes color singlet and color octet distributions
and is $\mu$ independent, whereas the $\mu$ dependence of $G$ and $K$ is
given by a renormalization group equation
\begin{align}
  \label{kernels-rg}
\gamma_K\bigl( \alpha_s(\mu) \bigr)
 &= - \frac{d K(\tvec{z}, \mu)}{d\log\mu}
    = \frac{d\ms G(x \zeta, \mu)}{d\log\mu}
\end{align}
and thus cancels in $G+K$.  Both $K$ and $\mathbf{M}$ are due to soft
gluon exchange and can be defined as vacuum matrix elements of Wilson line
operators, similar to those discussed in Section~\ref{sec:wilson}.  They
can only be calculated perturbatively if the transverse distances on which
they depend are sufficiently small.

The general solution of \eqref{general-cs-eq} can be written as
\begin{align}
  \label{general-cs-solution}
& \begin{pmatrix}
 \sing{F}(x_i, \tvec{z}_i, \tvec{y}; \zeta) \\[0.2em]
  \oct{F}(x_i, \tvec{z}_i, \tvec{y}; \zeta) \end{pmatrix}
 = e^{- S(x_1\zeta, \tvec{z}_1, \tvec{z}_2)
      - S(x_2\ms\zeta, \tvec{z}_1, \tvec{z}_2)}
\nonumber \\[0.2em]
&\qquad\quad \times
   e^{L\ms \mathbf{M}(\tvec{z}_1, \tvec{z}_2, \tvec{y})}\,
\begin{pmatrix}
  \sing{F}^{\mu_0}(x_i, \tvec{z}_i, \tvec{y}) \\[0.2em]
   \oct{F}^{\mu_0}(x_i, \tvec{z}_i, \tvec{y})
\end{pmatrix}
\end{align}
with
\begin{align}
  \label{sudakov-2}
& S(x \zeta, \tvec{z}_1, \tvec{z}_2)
 = {}- \frac{K(\tvec{z}_1, \mu_0) + K(\tvec{z}_2, \mu_0)}{2}\,
       \log\frac{x \zeta}{\mu_0}
\nonumber \\[0.2em]
&\quad + \int_{\mu_0}^{x \zeta} \frac{d\mu}{\mu}
  \biggl[ \gamma_K\bigl( \alpha_s(\mu) \bigr) \log\frac{x \zeta}{\mu}
        - G(\mu,\mu) \biggr]
\end{align}
and
\begin{align}
L &= \log\frac{\sqrt{x_1 x_2\rule{0pt}{1.5ex}}\, \zeta}{\mu_0} \;.
\end{align}
The scale $\mu_0$ specifies the initial condition of the differential
equation \eqref{general-cs-eq}, with a natural choice being $\mu_0 \propto
1 \big/ \sqrt{|\tvec{z}_1|\ms |\tvec{z}_2|}$.

The leading double logarithms of $\zeta/\mu_0$ in
\eqref{general-cs-solution} come from the second line in
\eqref{sudakov-2}, whereas terms involving $K$ and $\mathbf{M}$ only
contain single logarithms.  The Sudakov exponent $S$ also appears in the
solution of \eqref{cs-single} for single-quark distributions
\cite{Collins:1981uk},
\begin{align}
f(x,\tvec{z}; \zeta) &= e^{- S(x\zeta, \tvec{z}, \tvec{z})}\,
                        f^{\mu_0}(x,\tvec{z}) \,.
\end{align}
We thus obtain the important result that to double logarithmic accuracy
the Sudakov factor for a multiparton distribution is the product of the
Sudakov factors for single parton densities, both for color singlet and
color octet distributions.  A non-trivial cross talk between all partons,
and in particular a mixing between color singlet and octet distributions
occurs however at the level of single logarithms, which are known to be
important for phenomenology.  When the parton distributions $F(x_i,
\tvec{z}_i, \tvec{y}; \zeta)$ are inserted into the cross section formula
\eqref{X-sect-position}, the typical size of $|\tvec{z}_i|$ is $1
/|\tvec{q}{}_i|$ and $\zeta$ should be taken of order $Q$, so that
logarithms of $\zeta/\mu_0$ turn into logarithms of $Q/q_T$.

If all distances $\tvec{z}_i$ and $\tvec{y}$ are small, one can calculate
$K$ and $\mathbf{M}$ in perturbation theory.  We give explicit results in
\cite{Diehl:2011} and only mention some of their features here.  The
off-diagonal elements in the matrix $e^{L\ms \mathbf{M}}$ turn out to be
color suppressed, but only by $1/N$.  In the limit where $|\tvec{z}_i| \ll
|\tvec{y}|$ we find that the off-diagonal elements are additionally
suppressed by $|\tvec{z}_1|\ms |\tvec{z}_2| /\tvec{y}^2$ and that the
diagonal element for the color octet is smaller than the one for the color
singlet.  This results is a suppression of octet distributions by
\begin{align}
  \label{sudakov-suppr}
\frac{\oct{F}(x_i, \tvec{z}_i, \tvec{y};\zeta)}{%
     \sing{F}(x_i, \tvec{z}_i, \tvec{y};\zeta)} \sim
  \biggl( \frac{|\tvec{z}_1|\ms |\tvec{z}_2|}{\tvec{y}^2}
  \biggr)^\lambda 
\end{align}
with a power
\begin{align}
\lambda = \min\biggl(1, N\ms \frac{\alpha_s}{\pi}
  \log\frac{\sqrt{x_1 x_2\rule{0pt}{1.5ex}}\, \zeta}{\mu_0} \biggr) \,.
\end{align}
In the cross section one has $|\tvec{z}_i| \sim 1 /|\tvec{q}{}_i|$, so
that for large $q_T$ the factor \eqref{sudakov-suppr} disfavors color
octet distributions in a wide range of $\tvec{y}$.  To study
quantitatively the importance of color-octet suppression, one needs to
extend the perturbative result \eqref{sudakov-suppr} to the region of
non-perturbative distances $\tvec{y}$.  This has not been achieved yet.


\section{Conclusions}

We have studied several aspects of multiparton interactions in
hadron-hadron collisions.  Our theoretical framework is hard-scattering
factorization, which requires a large virtuality or momentum transfer in
each partonic scattering process but is valid in the full range of parton
momentum fractions.  A complementary approach, based on the high-energy
limit and using BFKL methods is discussed in
\cite{Ragazzon:1995cb,Braun:2000ua,Bartels:2005wa}.

The basic cross section formula for multiple interactions can be derived
at tree level using standard hard-scattering approximations and has an
intuitive geometrical interpretation in impact parameter space.  We have
shown that it can be formulated at the level of transverse-momentum
dependent multiparton distributions, which permits a description of the
transverse momenta of the particles produced in the hard scattering.  This
is particularly important because it is in transverse-momentum dependent
cross sections that multiparton interactions are not power suppressed
compared with single hard scattering.

The derivation of the tree-level formula for double hard scattering
exhibits nontrivial contributions from correlation and interference
effects.  They have partly been discussed earlier in the literature
\cite{Mekhfi:1985dv} but are not included in current phenomenological
models.  The polarizations of two partons can be correlated even in an
unpolarized proton, and we have shown that such correlations can
significantly affect both the overall rate and the angular distribution of
final-state particles in multiple interaction processes.  Two-parton
distributions have a nontrivial color structure since each parton can
carry a different color in the scattering amplitude and its complex
conjugate.  In addition, there are interference terms in fermion number
and flavor as shown in Fig.~\ref{fig:interference}.

To develop a reliable phenomenology, one needs information about the size
and kinematic dependences of two-parton distributions.  One can relate
them to generalized parton distributions for single partons, which are
experimentally accessible in exclusive scattering processes, but this
requires an approximation whose reliability we cannot quantify.
Nevertheless this relation offers some guidance, especially about the
interplay between longitudinal momentum and transverse position of
partons, as well as the size of two-parton distributions that are of
interference nature and hence do not have an intuitive probability
interpretation.  For transverse parton momenta above a few $\gev$ one can
compute transverse-momentum dependent distributions in terms of collinear
ones and thus has more predictive power from theory.  We find a tendency
for a simplified color structure in this regime, but quantitative
estimates remain to be done.  Finally, we have identified matrix elements
closely related to two-parton distributions that are suitable for
evaluation in lattice QCD.

To go from tree level to genuine factorization formulae, one must be able
to sum certain types of collinear and soft gluon exchanges into Wilson
lines.  We argue in \cite{Diehl:2011} that for double-scattering processes
producing color-singlet particles this can be achieved using the methods
that have been successfully applied to single Drell-Yan production
\cite{Collins:1981uk,Ji:2004wu,Collins:2007ph, Collins:2011}.  This also
permits the resummation of Sudakov logarithms, with important results
sketched in Section~\ref{sec:sudakov}.  For multiple jet production the
situation is unfortunately more complicated, as serious obstacles to
formulating transverse-momentum dependent factorization have been
identified even for single hard scattering \cite{Rogers:2010dm}.  The
production of two electroweak gauge bosons thus emerges as a channel where
the current perspectives for developing the theory look best, and where
different aspects of multiple interactions can hopefully be explored
experimentally at LHC.  For phenomenological estimates of $WW$ production
we refer to \cite{Kulesza:1999zh,Gaunt:2010pi}.

The cross section for double hard scattering involves an integral over the
transverse distance $\tvec{y}$ between the two partons emerging from each
hadron.  This includes the region of small $\tvec{y}$, where the two
partons can originate from the perturbative splitting of a single parton.
We find that this mechanism, which also affects the scale evolution of
multiparton distributions, leads to serious ultraviolet divergences in the
cross section and to a double counting problem between single and double
scattering.  To modify the definition of multiparton distributions and
possibly the cross section formulae is a prerequisite for putting the
theory on solid ground.


\section*{Acknowledgments}

We gratefully acknowledge discussions with J.~Bartels, J.~Bl\"umlein,
D.~Boer, V.~Braun, S.~Brodsky, F.~Ceccopieri, S.~Dawson, Ph.~H\"agler,
T.~Kasemets, P.~Kroll, Z.~Nagy, D.~Ostermeier, S.~Pl\"atzer, T.~Rogers,
G.~Salam, R.~Venugopalan and W.~Vogelsang.
A part of the calculations for this work was done using FORM
\cite{Vermaseren:2000nd}, and the figures were made with
JaxoDraw~\cite{Binosi:2003yf}.  This work was supported by BMBF
(06RY9191).


\end{document}